\documentclass[10pt,letterpaper,oneocolumn,prl,showpacs]{revtex4-1}

\usepackage[utf8]{inputenc} 
\usepackage[T1]{fontenc}    
\usepackage{hyperref}       
\usepackage{url}            
\usepackage{booktabs}       
\usepackage{amsfonts}       
\usepackage{nicefrac}       
\usepackage{microtype}      
\usepackage{lipsum}
\usepackage{graphicx}
\graphicspath{ {./images/} }

\usepackage{graphicx,color,textcomp}
\usepackage{mathrsfs,amsmath}   
\usepackage{psfrag,overpic,pst-all}
\usepackage{amssymb}

\begin{document}

\title{Nonlinear beam self-imaging and self-focusing dynamics in a GRIN multimode optical fiber: theory and experiments}

\author{Tobias Hansson$^{1}$}
\author{Alessandro Tonello$^{2}$}
\author{Tigran Mansuryan$^{2}$}
\author{Fabio Mangini$^{3}$}
\author{Mario Zitelli$^{4}$}
\author{Mario Ferraro$^{4}$}
\author{Alioune Niang$^{3}$}
\author{Rocco Crescenzi$^{4}$}
\author{Stefan Wabnitz$^{4,5}$}
\author{Vincent Couderc$^{2}$}

\affiliation{$^1$ Department of Physics, Chemistry and Biology, Link\"oping University, SE-581 83 Link\"oping, Sweden}
\affiliation{$^2$ Universit\'e de Limoges, XLIM, UMR CNRS 7252, 123 Avenue A. Thomas, 87060 Limoges, France}
\affiliation{$^3$  Dipartimento di Ingegneria dell'Informazione, University of Brescia, Via Branze 38, 25123 Brescia, Italy}
\affiliation{$^4$ Dipartimento di Ingegneria dell'Informazione, Elettronica e Telecomunicazioni, Sapienza University of Rome, Via Eudossiana 18, 00184 Rome, Italy}
\affiliation{$^5$Institute of Automation and Electrometry SB RAS, 1 ac. Koptyug ave., Novosibirsk 630090, Russia}

 




\begin{abstract}
Beam self-imaging in nonlinear graded-index multimode optical fibers is of interest for many applications, such as implementing a fast saturable absorber mechanism in fiber lasers via multimode interference. We obtain an exact solution for the nonlinear evolution of first and second order moments of a laser beam carried by a graded-index multimode fiber, predicting that the spatial self-imaging period does not vary with power. Whereas the amplitude of the oscillation of the beam width is power-dependent. We have experimentally studied the longitudinal evolution of beam self-imaging by means of femtosecond laser pulse propagation in both the anomalous and the normal dispersion regime of a standard telecom graded-index multimode optical fiber. Light scattering out of the fiber core via visible fluorescence emission and harmonic wave generation permits us to directly confirm that the self-imaging period is invariant with power. Spatial shift and splitting of the self-imaging process under the action of self-focusing are also emphasized.
\end{abstract}

\maketitle

\section{Introduction}

Nonlinear multimode optical fibers (MMFs) are an emerging research field, as they permit new ways for the control of spatial, temporal and spectral properties of ultrashort pulses of light  \cite{doi:10.1063/1.5119434}.
As a result, nonlinear MMFs are of interest for a diversity of optical technologies, e.g., for scaling-up the power of fiber lasers and supercontinuum light sources, for high-resolution biomedical imaging, micromachining, and high-power beam delivery, to name a few. MMFs also provide a simply accessible testbed for the study of complex physical phenomena \cite{Picozzi2015R30}. Among MMFs, graded-index fibers are of particular interest for nonlinear optics studies, since the reduced modal dispersion leads to relatively long interaction lengths among the fiber modes, even for pulses in the femtosecond regime \cite{6242367}.

Spatial self-imaging (SSI), first observed by Talbot in 1836 \cite{doi:10.1080/14786443608649032}, is a peculiar property of beam propagation in GRIN fibers. SSI leads to periodic oscillations along the fiber of the beam width and intensity.
In combination with the Kerr effect, SSI leads to a longitudinal periodic modulation of the refractive index of the fiber core, akin to a dynamic long-period grating. As a result, as recently discussed in a review paper by Agrawal \cite{AGRAWAL2019309}, SSI leads to many recently discovered nonlinear effects in GRIN MMFs, such as dispersive wave series emission from multimode femtosecond solitons \cite{Wright2015R31,PhysRevLett.115.223902}, geometric parametric instability of Continuous Wave (CW) beams in the normal dispersion regime \cite{PhysRevLett.116.183901}, and spatial beam self-cleaning phenomena \cite{Krupanatphotonics,WrightNP2016}.
In addition, SSI has been recently widely exploited for the mode-locking of fiber lasers, by exploiting multimode interference (MMI) resulting from SSI in a short piece of GRIN MMF with precisely controlled length, sandwiched between two singlemode fibers.

Because of SSI, any input field profile is periodically reproduced at equally spaced points $z_s$ along the GRIN fiber, such that

\begin{equation}\label{SI}
(\beta_n-\beta_1)z_s=\pi m_n,
\end{equation}

\noindent where $\beta_n$ and $\beta_1$ are the propagation constants of higher-order modes with index $n$ and of the fundamental mode, respectively, and $m_n$ is an integer. As a result, any input beam shape is reproduced at the fiber output, whenever its length is exactly equal to an integer multiple of $z_s$.
However, unless the shape of the input beam exactly matches the fundamental mode of the fiber, the input beam is spread in several propagating modes. As a result, the periodic reconstitution of the initial beam shape is obtained via the superposition of modes with different propagation constants, which affects the output beam divergence.
The presence of self-imaging in the MMF was indirectly put into evidence by Zhu et al. \cite{Zhu:08} by tuning the wavelength of the input signal, and measuring the transmission spectrum of the MMI structure.

Nazemosadat and Mafi proposed to use intensity-dependent differential phase shift among transverse modes, to obtain nonlinear MMI in a short length of GRIN MMF \cite{Nazemosadat:13}. According to their description, in a nonlinear MMF the self-imaging period should change to $z_l$ where

\begin{equation}\label{eq:SIN}
(\beta_n(I)-\beta_1(I))z_I=\pi m_n.
\end{equation}

In a realistic situation, an input laser beam coupled to a GRIN MMF excites several modes, each of them carrying a different amount of intensity along the fiber.
Therefore the simple description of Eq.~(\ref{eq:SIN}) should be substituted by the solution of nonlinear coupled mode equations including self- and cross-phase modulation, as well as four-wave mixing terms \cite{Poletti:08}. Equivalently, one may use a two-dimensional nonlinear Schr\"odinger equation (2D-NLSE) ~\cite{Krupanatphotonics}. 

By sandwiching the MMF between two singlemode fibers (SMS structure), it has been experimentally observed that nonlinear MMI acts as a fast saturable absorber (SA) mechanism which permits to obtain mode-locking in high pulse-energy fiber lasers. Specifically, low power signals are strongly attenuated when propagating through the SMS structure, while high intensity pulses experience a relatively high transmission. Nonlinear MMI-based SAs permit to operate at much higher pulse energies and peak powers than other SA mechanisms, thanks to their high damage threshold, low cost, simple structure, and mechanical robustness.
Nonlinear MMI in a GRIN MMF also permits spectral filtering, a necessary property for fiber laser mode-locking in the normal dispersion regime.

Several different variants of the technique of fiber laser mode-locking based on nonlinear MMI have been demonstrated in recent years
\cite{Fu:16,8093610,Li:17,Yang:18,Wang:18a,Tegin:18,Zhao_2018,Zhao:2018,Chen:19}.
In order to test the nonlinear transmission of the SMS structure, the MMF was stretched, and the SMS transmission was measured as a function of the stretching length, for different input intensities \cite{Wang:18a}. It was found that, at relatively high intensities, the self-imaging induced beam oscillations occur with much smaller amplitude, and with average (i.e., over one self-imaging period) transmission that is significantly increased ($\simeq 75\%$) with respect to the low intensity case ($\simeq 68\%$). This suggests that nonlinear mode coupling and the resulting beam reshaping could be a mode-locking mechanism, rather than the nonlinear variation of the self-imaging period.
Indeed, experiments show that the excitation of high-order modes (HOMs) in the GRIN fiber is a necessary condition to obtain saturable absorber action \cite{Li:17,8093610}. A study of the impact of the GRIN core diameter (ranging from 20 to 62.5 $\mu m$) revealed that larger diameters (hence larger proportions of excited HOMs) increase the output laser power.
Another mode-locking mechanism could result from the longitudinal translation of the self-imaging process, because of self-focusing effects in the GRIN MMF.


Already back in 1992, Karlsson et al. have theoretically studied, by means of the variational method, the dynamics of self-imaging in a GRIN MMF under the combined action of diffraction, nonlinearity, and parabolic index profile \cite{Karlsson:92}. They derived an approximate analytical solution for a multimode beam, which permit to evaluate the evolution of the beam width and its longitudinal phase delay. In this work, we derive an exact solution for the nonlinear evolution of first and second order moments of a laser beam carried by a GRIN multimode fiber. Our theoretical analysis does not require, unlike methods based on the variational approach, that the beam maintains a specific shape (e.g., gaussian) throughout its propagation in the MMF. This permits us to show, in full generality, and in contrast with current understanding of nonlinear MMI based SAs, that the SSI period does not vary with power. Whereas the amplitude of the beam width oscillations is power-dependent, which is still sufficient to enable a power-dependent transmission for the SA. We also derive the power threshold for catastrophic self-focusing in the GRIN MMF.

Although SSI is a well-known process, it remains difficult to directly prove its existence in MMFs. Here we show that it is possible to work around this problem, by experimentally recording the local nonlinear parametric conversion obtained in a non-collinear geometry, i.e., by using Cherenkov phase matching \cite{Cherenkov:1934}.  As we shall see, the periodic emission of a second harmonic wave, obtained at the core-cladding interfaces, accompanied by multi-photon absorption leading to wideband fluorescence in the blue spectral domain, provides a clear evidence of the periodic evolution of the power density in the MMF. We could highlight these processes, we believe for the first time, by using femtosecond laser pulse excitation both in the anomalous and in the normal dispersion regime. This permits us to obtain sufficiently high peak powers, in the multi-MW range, at the points of minimum beam waist of the SSI process.

\section{Theory}
\label{sec:TH}

\subsection{Model}

We consider beam propagation in a multimode optical fiber with a parabolic index profile. The beam dynamics is assumed to be described by a 2D-NLSE with an instantaneous Kerr nonlinearity of the form
\begin{equation}
  \frac{\partial A}{\partial z} - i\frac{1}{2k_0}\left(\frac{\partial^2 A}{\partial x^2}+\frac{\partial^2 A}{\partial y^2}\right) + i\frac{k_0\Delta}{\rho^2}(x^2+y^2)A = i\frac{k_0n_2}{n_{co}}|A|^2A,
  \label{eq:NLS}
\end{equation}
where $k_0 = \omega_0n_{co}/c$ is the wavenumber, $\rho$ is the core radius, $\Delta = (n_{co}^2-n_{cl}^2)/2n_{co}^2$ is the relative index difference, and $n_{co}$ ($n_{cl}$) is the refractive index of the fiber core (cladding), respectively. Here we neglect dispersion and Raman scattering and assume that the index profile has an infinite extent (i.e.~it does not truncate when the cladding index is reached). 

\subsection{Moments of the beam}

Now, we show that it is possible to obtain an exact, analytical solution for the moments of the field, whose evolution is described by the 2D-NLSE~(\ref{eq:NLS}). In particular, we derive a closed system of equations for the various moments.
The moments may conveniently be obtained by considering them as expectation values of observables and using operator methods from quantum mechanics. With this machinery the evolution equations for the various moments can be found in a few lines of calculation using commutator algebra. We introduce a linear Hamiltonian operator
\begin{equation}
  \hat{H} = \frac{1}{2k_0}\left(\hat{p}_x^2+\hat{p}_y^2\right) + \beta\left(\hat{x}^2+\hat{y}^2\right)-\gamma \hat{I}
\end{equation}
where $\beta = k_0\Delta/\rho^2$, $\gamma = k_0n_2/n_{co}$, $\hat{x}$ and $\hat{y}$ are transverse position operators, and the nonlinear term $\hat{I} = |A(x,y)|^2$ is treated as a potential term that should be determined self-consistently. Here we have also made use of the momentum operators $\hat{p}_x = -i\partial/\partial x$ and $\hat{p}_y = -i\partial/\partial y$. With this pseudo-Hamiltonian we can write Eq.~(\ref{eq:NLS})  as
\begin{equation}
  i\frac{\partial A}{\partial z} = \hat{H}A.
\end{equation}
Following the procedure described in the Appendix A, we find the sinusoidal solutions for the first-order moments
\begin{equation}
  \langle x\rangle = a_x\cos{\left(\sqrt{\frac{2\beta}{k_0}}z+b_x\right)}, \qquad
  \langle y\rangle = a_y\cos{\left(\sqrt{\frac{2\beta}{k_0}}z+b_y\right)},
    \label{eq:sinus}
    \end{equation}
where the oscillation period depends on the index parameters, and $a_{x,y}$, $b_{x,y}$ are integration constants that depend on the initial conditions.
Moreover, the radial moment $\langle x^2+y^2\rangle = \langle x^2\rangle + \langle y^2\rangle$, associated with the rms-width for a centered beam, satisfies the closed equation
\begin{equation}
  \frac{d^2\langle x^2+y^2\rangle}{dz^2} = \frac{2}{k_0}\langle H_0\rangle - \frac{8\beta}{k_0}\langle x^2+y^2\rangle.
  \label{eq:rms_eq}
\end{equation}
where the Hamiltonian invariant $\langle H_0\rangle$ is defined in Eq.~(\ref{eq:H0}). This equation has a solution when $\langle H_0\rangle > 0$ that describes sinusoidal oscillations around a constant average value, viz.
\begin{equation}
  \langle x^2+y^2\rangle (z) = \frac{1}{4\beta}\langle H_0\rangle + a_r\cos\left(\sqrt{\frac{8\beta}{k_0}}z+b_r\right),
  \label{eq:rms}
\end{equation}
where $a_r$ and $b_r$ are integration constants that determine the amplitude and phase of the oscillations. These can be found from the initial beam profile entering the fiber. Whenever
the input beam is not in the process of changing, i.e.~$[d\langle x^2+y^2\rangle/dz]_0 = 0$, one obtains
$a_r = \langle x^2+y^2\rangle_0 - \langle H_0\rangle/4\beta$ and $b_r = 0$ . The oscillation amplitude is consequently a function of the input beam width, while the SSI period is determined by the fiber parameters only, and it remains equal to its linear value $z_s=\pi \rho/\sqrt{2\Delta}$. It is also seen that the oscillation period of the first-order moments is twice that of the rms-width.

The first integral of Eq.~(\ref{eq:rms_eq}) represents a conservation law, $M^2 - \gamma k_0\langle x^2+y^2\rangle\langle I\rangle = \textrm{const.}$, for the beam quality factor $M^2 = \langle x^2+y^2\rangle\langle p_x^2+p_y^2\rangle-\frac{1}{4}\left(\langle xp_x+p_xx\rangle+\langle yp_y+p_yy\rangle\right)^2$, cf.~~\cite{Dragoman:96}. Furthermore, assuming $b_r = 0$, we may rewrite Eq.~(\ref{eq:rms}) by using the beam compression parameter (C-parameter) that was introduced by Karlsson et al.~\cite{Karlsson:92}
\begin{equation}
  \langle x^2+y^2\rangle (z) = \frac{\langle x^2+y^2\rangle_0}{2}\left[\left(1+C\right)+\left(1-C\right)\cos\left(\sqrt{\frac{8\beta}{k_0}}z\right)\right], \qquad C = \frac{1}{2\beta}\frac{\langle H_0\rangle}{\langle x^2+y^2\rangle_0}-1
  \label{eq:Crms}
\end{equation}
This parameter measures the relative importance of diffractive broadening and nonlinear self-focusing. The radial moment is seen to be stationary for $C = 1$, while it becomes negative for $C < 0$ which corresponds to the condition for collapse. The critical collapse power is reduced in a GRIN fiber with respect to a homogeneous bulk material, because of the guiding refractive index profile. In the latter case the Hamiltonian must be negative for collapse to occur for a beam without an initial phase front curvature \cite{HANSSON20113422}.

\subsection{Vortex beam}

\begin{figure}[ht]
  \centering
  \includegraphics[width=.75\linewidth]{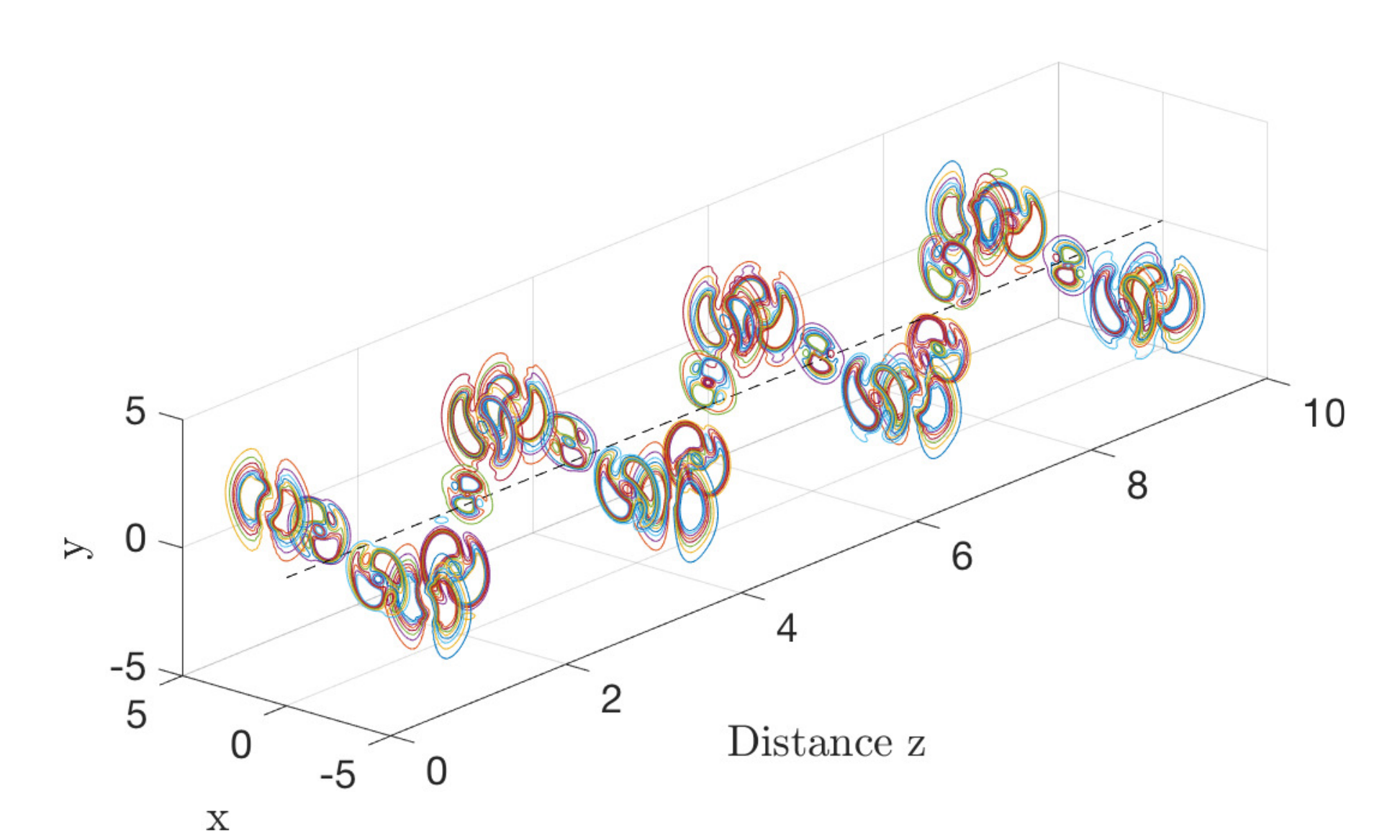}
     \includegraphics[width=.45\linewidth]{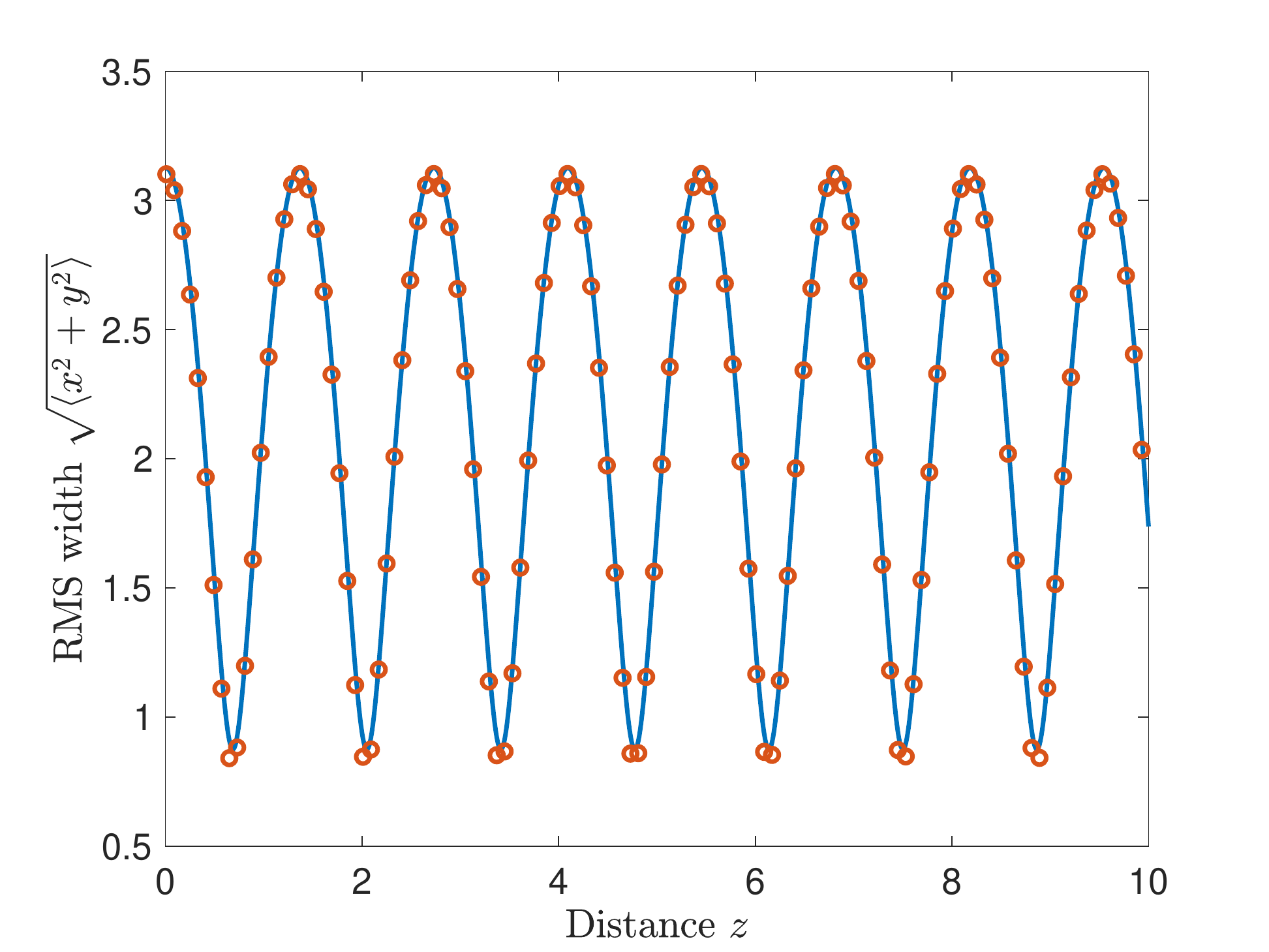}
\caption{Top panel: Contour slices showing simulated propagation of a complex beam with a vortex charge. The initial beam profile has been displaced from the center of the fiber (dashed line) and displays periodic oscillations in both position and rms-width; Bottom panel: Evolution of the moment corresponding to the non-centered rms-width}
  \label{fig_vortex}
  \end{figure}

A numerical example of a propagating beam with a non-trivial beam profile is shown in Fig.~\ref{fig_vortex}, together with the corresponding evolution of the associated moment. The normalized simulation parameters are $k_0 = 1.3, \Delta = 1.7, \rho = 0.8, n_{co} = 2.0$ and $n_2 = 1.5$, and the initial field is given by the function $A = A_0~\textrm{sech}(x/a_x)\textrm{sech}(y/a_y)\sin(x)e^{i2\textrm{tan}^{-1}(y/x)}$ with coefficients $a_x = 0.81, a_y = 1.01$ and $A_0 = 0.56$ that has been shifted to the position $(x_0,y_0) = (1.29, 2.47)$. The simulated moment (red dots) is found to be in excellent agreement with the analytical expressions that is shown as a blue curve in Fig.~\ref{fig_vortex} (any discrepancy is only due to numerical inaccuracy).

\subsection{Gaussian beam}

\begin{figure}[ht]
  \centering
  \includegraphics[scale=.5]{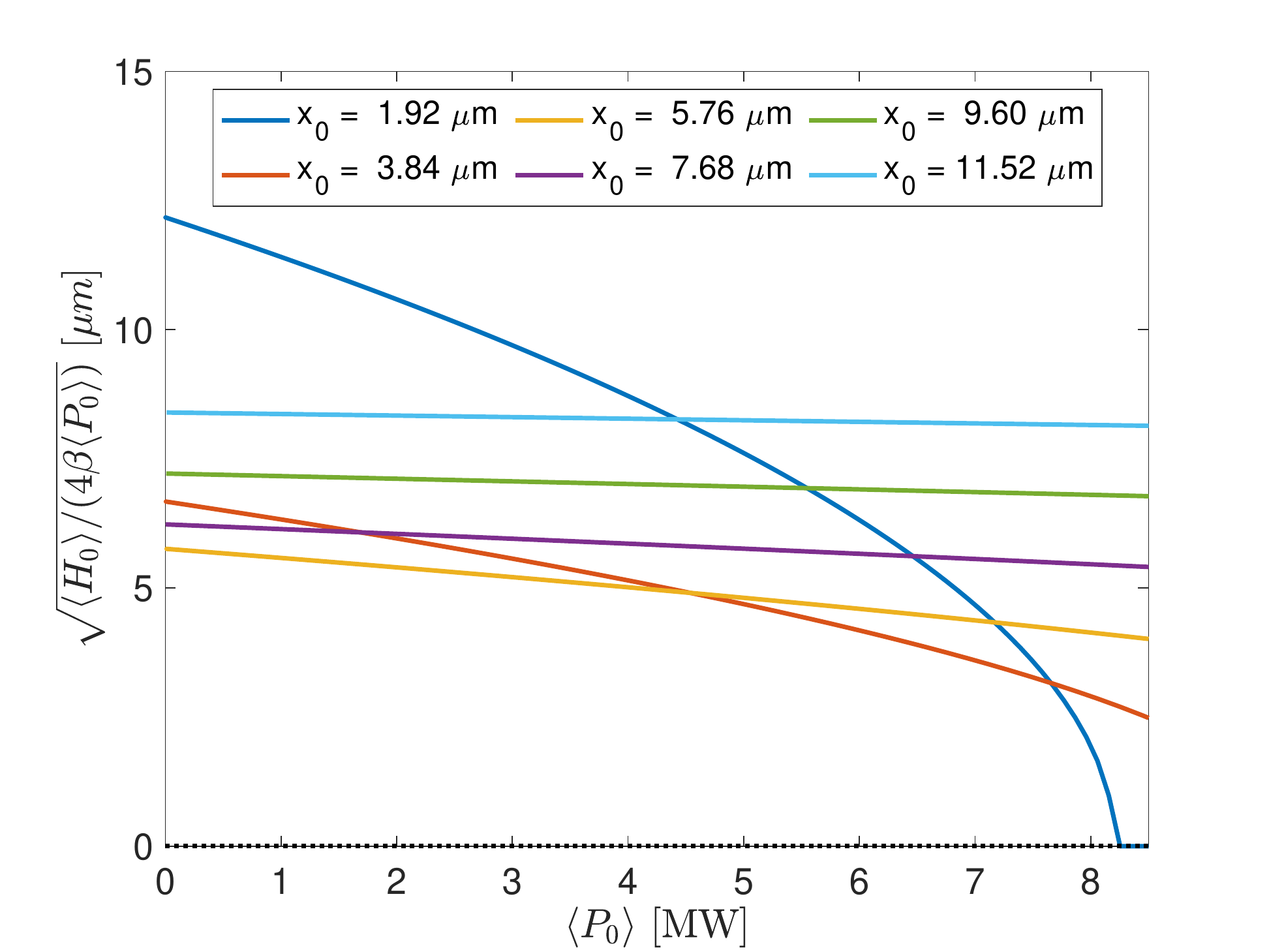}
  \caption{Variation of average beam width with power, for unchirped Gaussian initial conditions.}
  \label{fig_beamw}
  \end{figure}

It is also evident from Eq.~(\ref{eq:rms}) that the constant average value for the rms width depends on the Hamiltonian invariant that is determined by the initial conditions. Assuming a Gaussian initial beam profile of the form
\begin{equation}
  A(x,y,0) = A_0\exp\left[-\frac{1}{2}\left(\frac{x}{x_0}\right)^2-\frac{1}{2}\left(\frac{y}{y_0}\right)^2\right],
\end{equation}
the power and Hamiltonian invariants evaluate to
\begin{equation}
  \langle P_0\rangle = \pi|A_0|^2x_0y_0, \qquad \langle H_0\rangle = \frac{\langle P_0\rangle}{2}\left[\frac{1}{k_0}\left(\frac{1}{x_0^2}+\frac{1}{y_0^2}\right)+2\beta(x_0^2+y_0^2)-\frac{\gamma \langle P_0\rangle}{\pi x_0y_0}\right].
\end{equation}
The ($1/e$) beam width is found by calculating the radial moment $\sqrt{\langle x^2+y^2\rangle/\langle P_0\rangle} = \sqrt{(x_0^2+y_0^2)/2}$, which reduces to $x_0 = y_0$ for a symmetric beam. It is seen that the magnitude of the Hamiltonian is reduced by the nonlinear term for a focusing nonlinearity. This implies that the average value of the radial beam width will shrink as the power is increased. In Fig.~\ref{fig_beamw} we illustrate the power-induced variation of the average beam width, i.e.~$w = \sqrt{\langle H_0\rangle/4\beta\langle P_0\rangle}$, for different values of the input width $x_0$. As can be seen, the average beam width decreases as the power $\langle P_0\rangle$ grows larger: the narrower the input beam size $x_0$, the faster the nonlinear decrease of the beam width with power.

When using fiber parameters from the numerical simulations reported in Ref.~ \cite{Krupanatphotonics}, and assuming a fixed input beam diameter of $40~\mu m$ (FWHM) so that $x_0 = y_0 = 24.0~\mu m$, one obtains that associated power-induced variation of the average width for the radial moment $w$ occurs on a GW scale (and $w$ shrinks down to zero for $\langle P_0\rangle=2.5$ GW). This means that the nonlinear reduction of the Hamiltonian is not the root cause of spatial beam cleaning. Note that, in this case, the average radial beam width approaches the value $w = 17.0~\mu m$ (c.f.~core radius $\rho = 26~\mu m$) in the low power limit. This can be compared with the radial width of the fundamental linear eigenmode solution
\begin{equation}
  A(x,y,z) =  \sqrt{\frac{\alpha}{\pi}}\exp\left[-\frac{1}{2}\alpha(x^2+y^2)-i\frac{1}{k_0}\alpha z\right],  \qquad \alpha = \sqrt{2k_0\beta},
\end{equation}
which has $w_f=\sqrt{\langle x^2+y^2\rangle/\langle P_0\rangle} = 1/\sqrt{\alpha} \approx 5.76~\mu m$ or $9.59~\mu m$ FWHM (see corresponding orange curve on Fig.~\ref{fig_beamw}). Note that, for an input beam perfectly matching the fundamental mode shape, the average beam width shrinks until it reaches zero when $\langle P_0\rangle \approx 16.55~\textrm{MW}$. However, the beam will not be stationary in the GRIN profile, and oscillations of the beam width will cause it to collapse at a lower critical power. The moment theory is however exact under the assumption of a non-truncated index profile, which suggests that a cleaned beam should still have an overall rms-width, although perhaps with a skewness, that is in agreement with the theory.

\begin{figure}[ht]
\includegraphics[scale=.5]{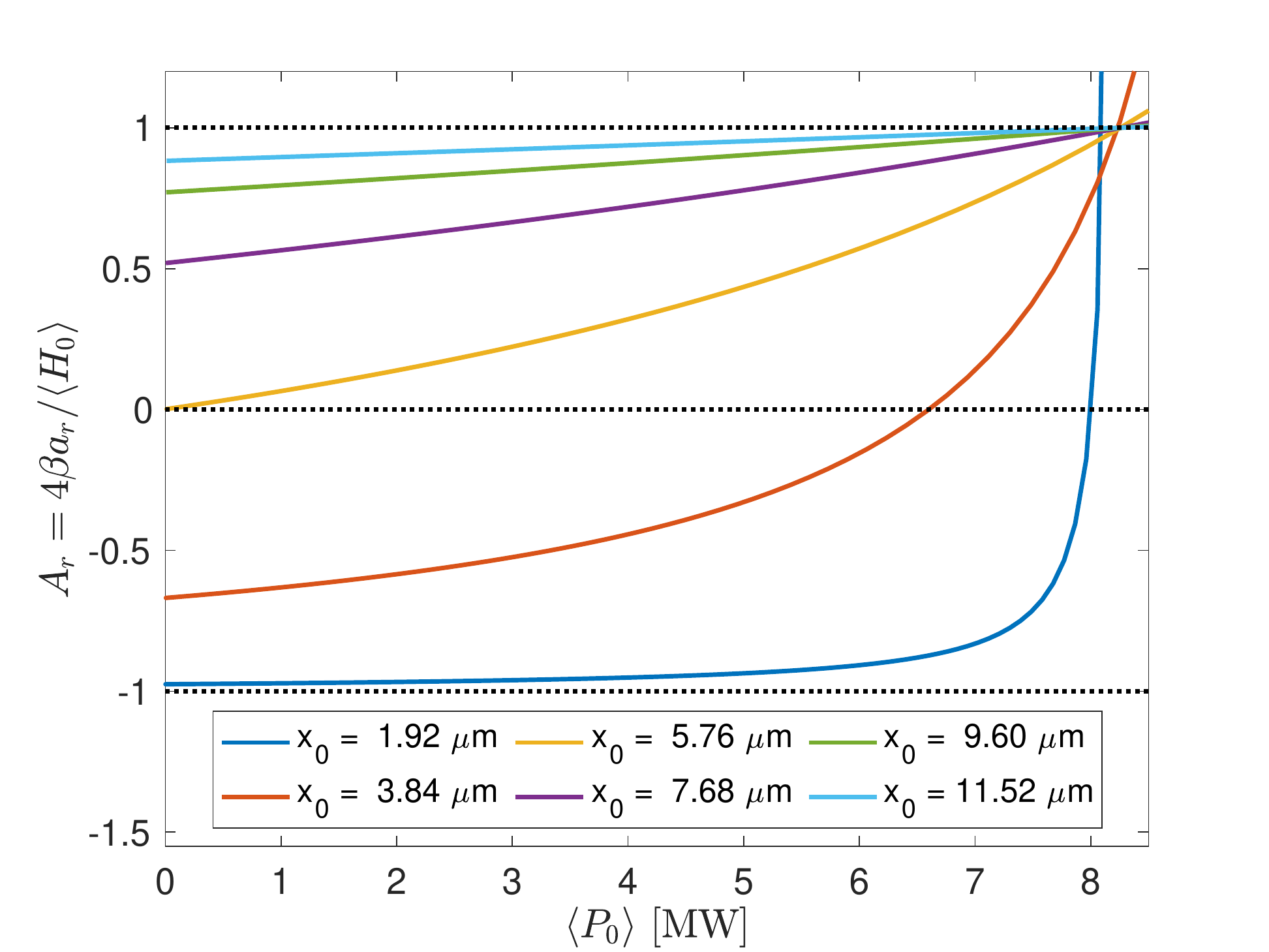}
 \centering
      \caption{Variation of beam oscillation amplitude with power, for unchirped Gaussian initial conditions.}
    \label{fig_ampl}
    \end{figure}

To better highlight the nonlinear dependence of the solution for the second-order radial moment Eq.~(\ref{eq:rms}), we plot in Fig.~\ref{fig_ampl} the input beam power $\langle P_0\rangle$ dependence of the beam width oscillation amplitude $A_r = 4\beta a_r/\langle H_0\rangle$, for different values of the the input radial width $w$.  As can be seen, if the initial width $w=w_f$, the oscillation amplitude is equal to zero in the linear limit, and it increases with power until it reaches unity for
$\langle P_0\rangle \simeq 8.3 MW$. This means that the beam width shrinks to zero, which corresponds to a beam collapse condition. In fact, the C-parameter
\begin{equation}
  C = \frac{1}{2\beta k_0x_0^2y_0^2}\left(1-\frac{\gamma k_0\langle P_0\rangle x_0y_0}{\pi(x_0^2+y_0^2)}\right)
\end{equation}
shows that, for a symmetric Gaussian with flat initial phase, the critical collapse power $\langle P_c\rangle = 2\pi/(\gamma k_0)=\lambda^2/(2\pi n_{co}n_2)$ is independent of the beam width.
For a beam with a nonvanishing initial phase-front curvature, our approach also permit to obtain the condition for critical collapse to occur. It is interesting to point out that the collapse power depends on the incidence angle of the beam into the GRIN MMF. Details about the analytical description of SSI beam evolution in the most general case is provided in the Appendix.

Fig.~\ref{fig_ampl} also shows that, in the low-power limit, the oscillation amplitude is positive (negative) when the input beam width is larger (smaller) than the width of the fundamental mode $w_f$. A positive (negative) value of $A_r$ means that the beam width is decreasing (increasing) along the fiber with respect to the input value, until it experiences a maximum beam compression (widening), before returning back to the input width after one period. At sufficiently high powers, self-focusing leads to a narrowing beam width in all cases.

The generality of the moments method also permits to analytically study the SSI dynamics for beams with non-Gaussian transverse profile. We have thus generalized the analytcal description to the case of a super-Gaussian initial beam profile: corresponding results are reported in the the Appendix.

\section{Experiments}

In order to confirm the theoretically predicted invariance of the SSI spatial period $z_s$ with respect to the input power, we carried out an experimental study of the dynamics of SSI in the nonlinear regime of pulse propagation in a GRIN MMF. By using femtosecond input pulses with peak powers up to the MW power range, we could directly visualize the high intensity points inside the MMF, thanks to the associated side scattering of visible higher-harmonics and fluorescence light.

\begin{figure}[ht]
  \centering
  \includegraphics[scale=.45]{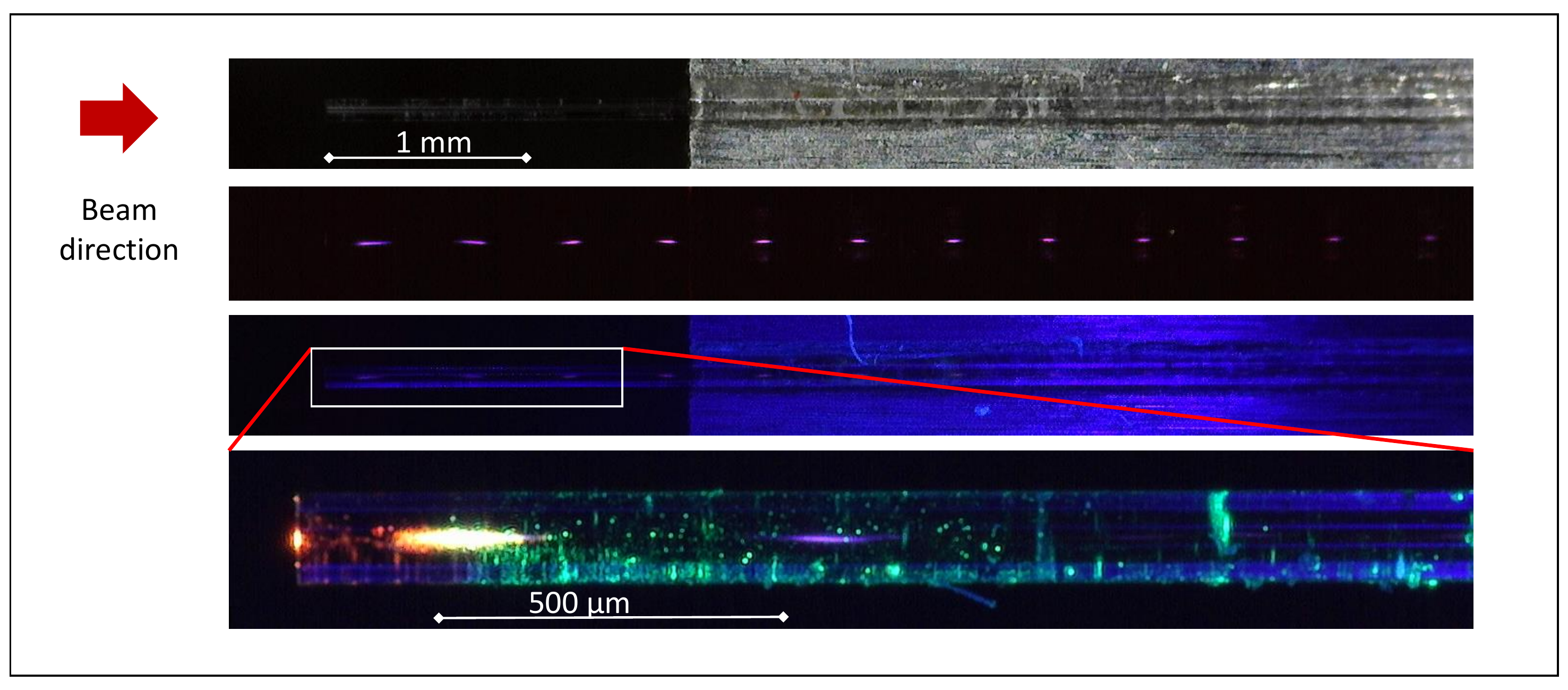}
  \caption{Side scattering imaging of periodic self-imaging in 5 cm of GRIN MMF. From the top, we show the digital microscope picture, fluorescence at $\sim$7 MW of input peak power, the side scattering of violet radiation, and its zoom.}
    \label{immagini_iniziali}
    \end{figure}

\subsection{Anomalous dispersion}

We performed two sets of measurements, with two different laser sources. First, we used an ultra-short femtosecond laser
system, involving a hybrid optical parametric amplifier (OPA) of white-light continuum, pumped by a femtosecond
Yb-based laser, generating 120 fs pulses at 1550 nm, with 25 kHz repetition rate. The input laser beam was focused by a 30 mm focal lens, corresponding to an input beam diameter ($1/e^2$) of 18 $\mu$m, into a 5 cm long multimode standard 50/125 GRIN fiber, with relative index difference $\Delta=0.0102$.
    
\begin{figure}[htb!]
  \centering
    \includegraphics[scale=.15]{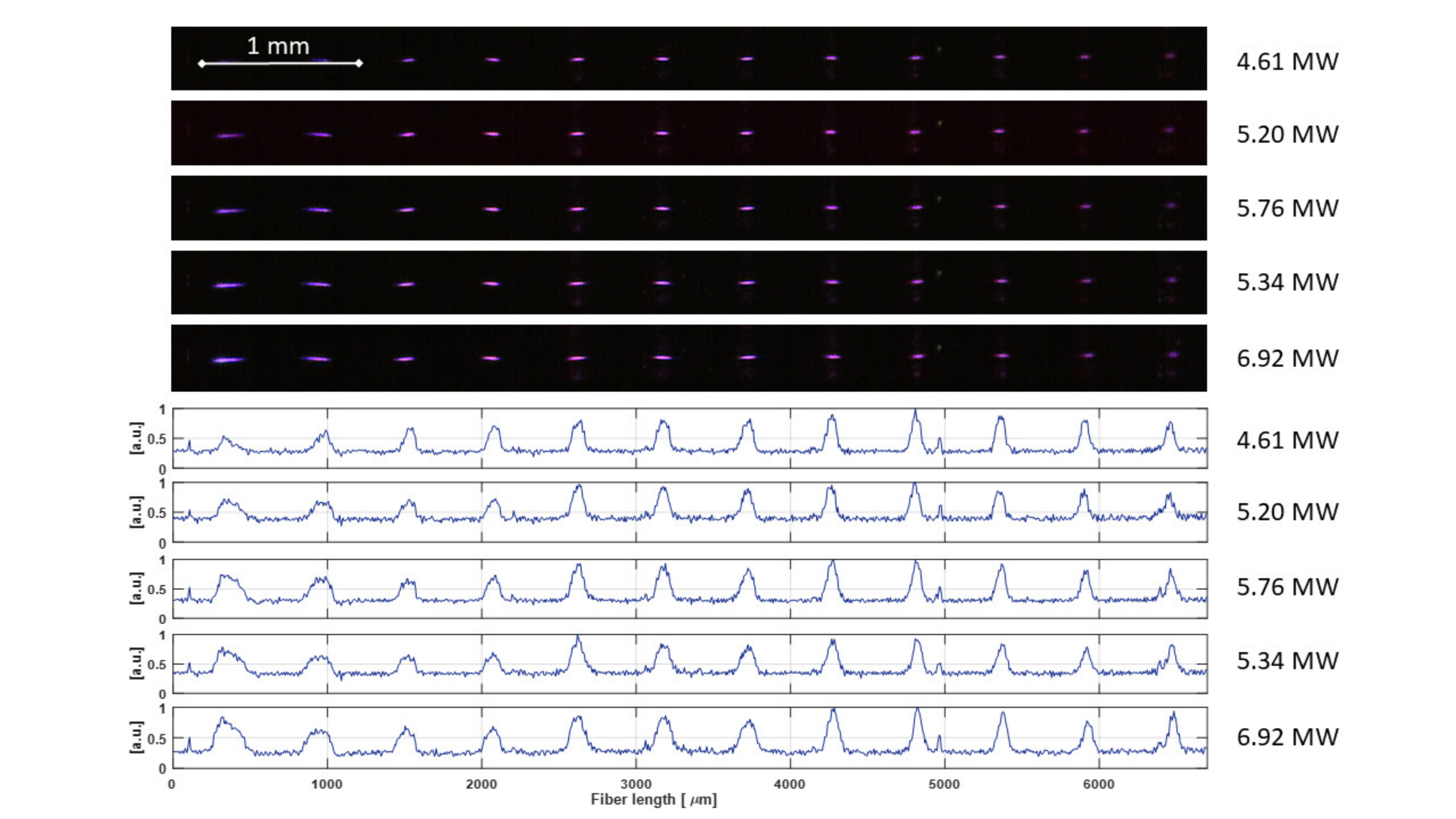}
  \caption{Experimental mesurements of periodic SSI in 5 cm of GRIN MMF, for 5 different input peak power values. Top panels: digital microscope pictures. Bottom panels: relative 1D plot of the fluorescence intensity.}
    \label{analisi_immagini_new}
    \end{figure}

Figures~\ref{immagini_iniziali} and ~\ref{analisi_immagini_new} reveal the presence of multiple peaks of scattered light, corresponding to the nodes of maximum beam compression in the course of SSI, and highest intensities. This permits us to directly monitor, as shown in Figure~\ref{analisi_immagini_new}, the dependence of SSI as a function of the input peak power of the injected pulses.
Here the input peak power that is coupled into the MMF is increased from 2.3 MW up to 7 MW.

\begin{figure}[htb!]
  \includegraphics[scale=.33]{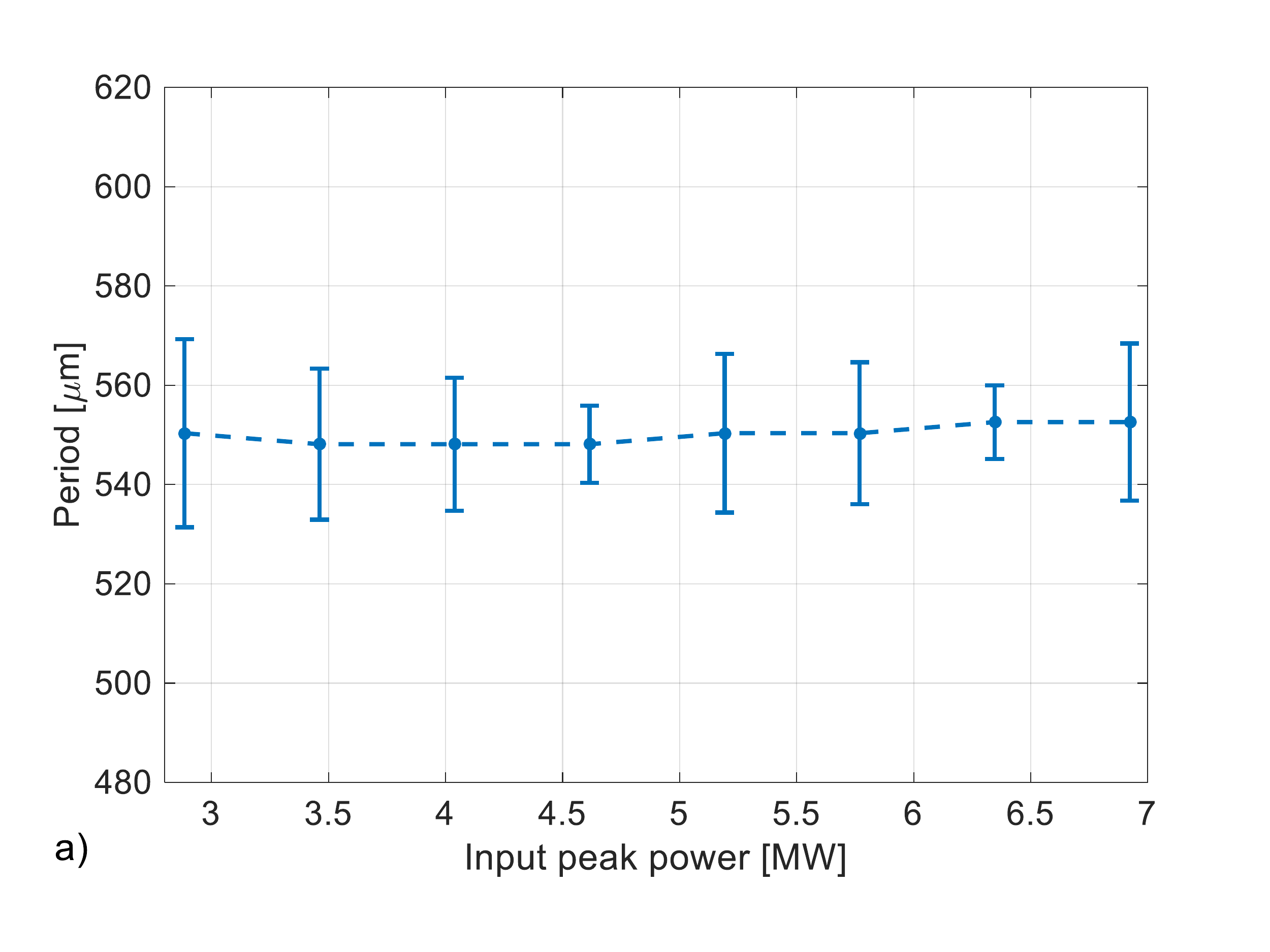} 
   \includegraphics[scale=.33]{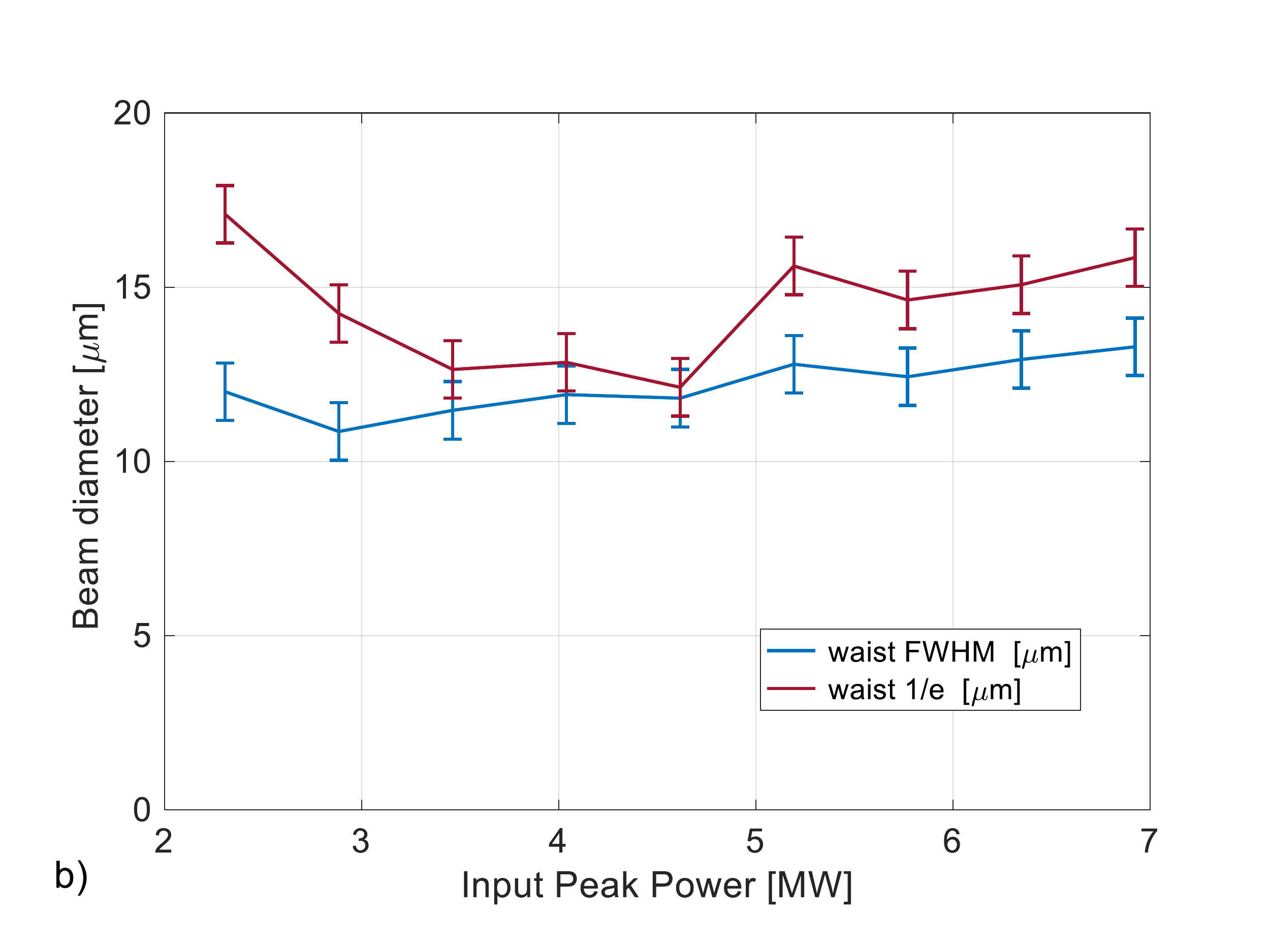}  
     \includegraphics[scale=.13]{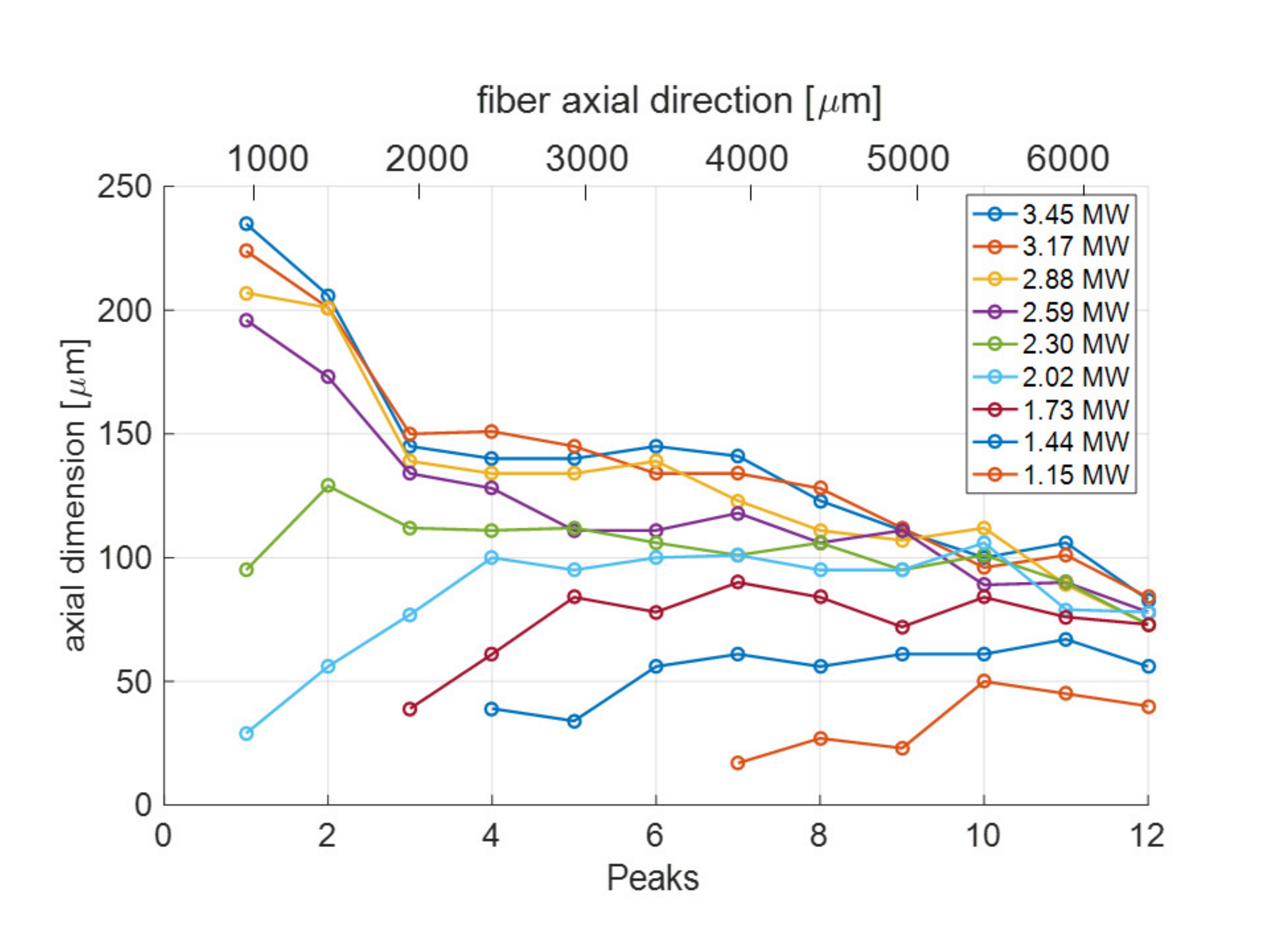}
     \centering
  \caption{Experimental measurements of power dependence of (a) self-imaging period; (b) beam transverse and (c) axial dimension: vertical lines represent the variance of the measurements for each input power.}
    \label{analisi_immagini_old2}
    \end{figure}

Figure~\ref{immagini_iniziali} shows, in the top panel, the picture taken by a digital microscope. The second panel from the top shows that blue fuorescence is periodically scattered from the side of the MMF. The bottom two panels in Figure~\ref{immagini_iniziali} show that green light is also scattered by the defects of the fiber cladding.

As can be seen in Figure ~\ref{analisi_immagini_new}, the SSI period remains unchanged, and remarkably close to the theoretically predicted value (i.e., $z_s=\pi \rho /\sqrt{2\Delta} \approx 550 \mu m$).
The input pulses generate high-order multimode solitons in the fiber. These solitons undergo, after a propagation distance of about 10 cm, fission into several fundamental solitons under the action of Raman soliton self-frequency shift and higher-order dispersion. Here we limit ourselves to consider the regime of multisoliton propagation over the first few cm of GRIN MMF, that is before that soliton fission takes place. The dynamics of multimode Raman solitons generated by the fission process will be discussed in a separate publication. However, it is important to point out that the observed processes of multiphoton absorption-induced side-scattering of visible fluorescence lead to significant nonlinear losses, which induce intensity clamping at the output of the GRIN MMF.

   
\begin{figure}[htb!]
 \centering
  \includegraphics[scale=.4]{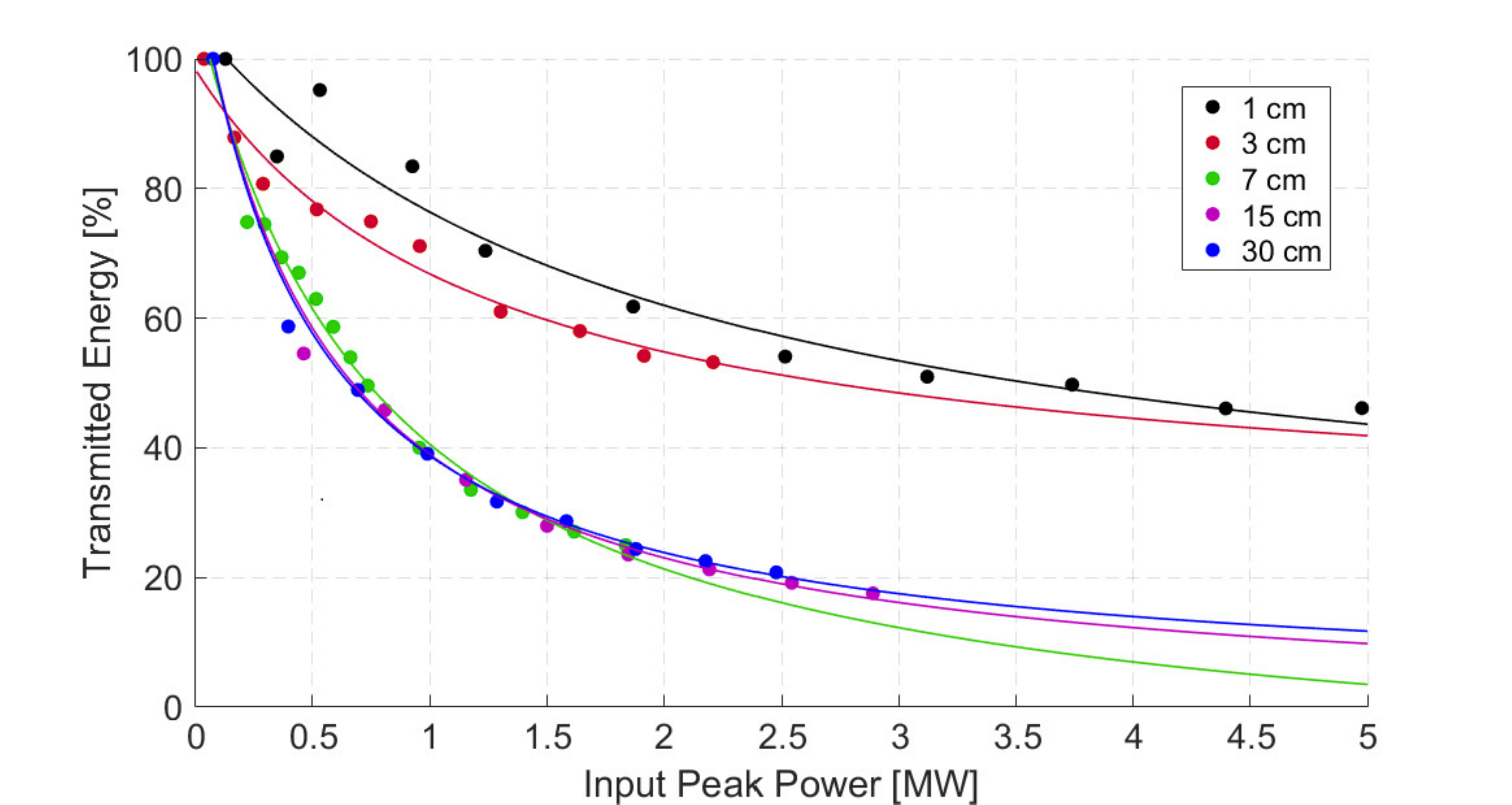} 
  \caption{Experimental measurement of nonlinear transmission, for different lengths of the GRIN fiber. Dots: experimental points; curves are provided as a guide to the eye.}
    \label{cut}
    \end{figure}
  
The invariance of the SSI period with power is well illustrated by Figure~\ref{analisi_immagini_old2}(a), that shows the input power dependence of the spatial frequency intensity spectrum of the scattered fluorescence. These results confirm that the SSI period remains a constant (within experimental measurement errors) even at the highest input peak powers.

We also measured the power-dependence of the beam size at the points of maximum beam compression, corresponding to the bright spots in Figure ~\ref{analisi_immagini_new}.
Figure~\ref{analisi_immagini_old2}(b) reports the dimension of the beam in the transverse section of the fiber: as can be seen the beam size does not show a significant dependence on input  power. This confirms the hypothesis of Section \ref{sec:TH} that the ansatz for the transverse mode profile is maintained along the nonlinear beam propagation.
On the other hand, Figure~\ref{analisi_immagini_old2}(c) shows that the beam size in the axial dimension exhibits a power-dependence. 

 Specifically, Figure~\ref{analisi_immagini_old2}(c) shows that, for input peak powers below (above) 4 MW, the beam dimension in the axial direction increases (decreases) along the fiber. The longitudinal dependence of the beam size in the axial direction could be related with the presence of the significant nonlinear losses, which occur over the first few centimeters of the fiber. To put into evidence the nonlinear transmission properties of the GRIN MMF subject to input peak powers just below the critical value for collapse, we reported in Figure~\ref{cut} the values of the fraction of input coupled energy that emerges from different lengths of the fiber, as a function of the input peak power.
 As can be seen, from 1 cm of fiber the transmission drops below 50\% at the highest peak powers close to 5 MW. Whereas for lengths above 7 cm, the transmission drops below 20\% for powers approaching 3 MW. Note that the different curves for fiber lengths above 7 cm tend to overlap, indicating that most of the nonlinear loss occurs over the first few centimeters of the fiber.

  
  \subsection{Normal dispersion}

  \begin{figure}[htb!]
  \centering
  \includegraphics[scale=.6]{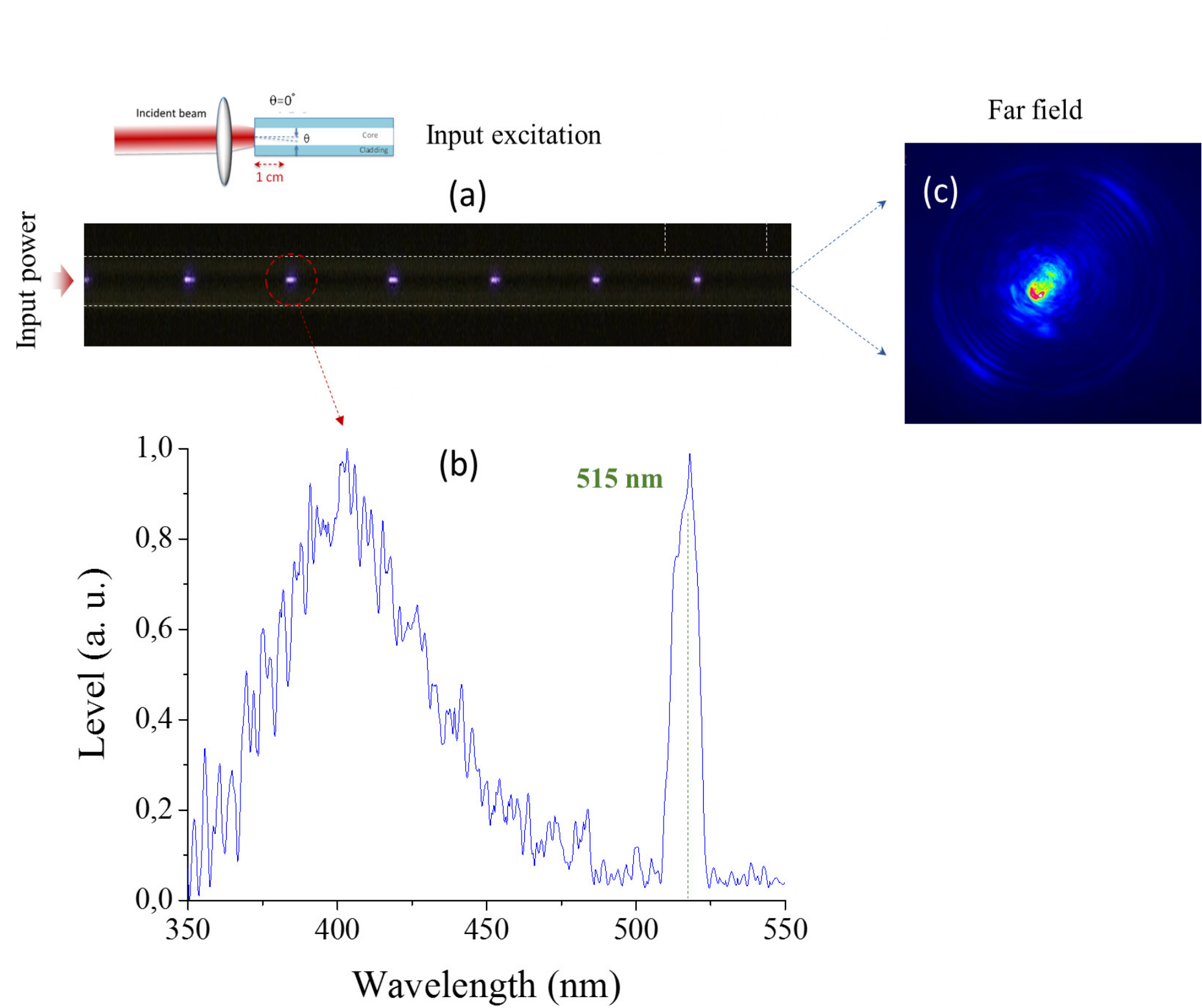}
  \caption{Experimental results on the non-collinear frequency conversion obtained in a 50/125 GRIN fiber by using a 250 fs laser source at 1030 nm; (a) schematic representation of the coupling conditions and side image of the fiber, showing periodic emission of visible light; (b) Spectrum of the emitted light, (c) far field image of the output beam (fiber length: 1 cm, peak power: 2MW).}
    \label{Fig8.eps}
    \end{figure}
  
We also carried out a series of measurements in the normal dispersion regime of the fiber, by using a fiber laser source at 1030 nm, generating 250 fs pulses with a 30 kHz repetition rate. Here we injected the pulses in a relatively short, 1 cm section of a 50/125 $\mu$m GRIN MMF. We obtained at the nodes (that is, at the points of minimum beam waist and maximum peak intensity) of SSI a sufficient intensity to trigger, in addition to multi-photon fluorescence, also non-collinear (i.e., using Cherenkov phase matching \cite{Cherenkov:1934}) second-harmonic generation (SHG), both of which are scattered outside the fiber cladding (see Figure~\ref{Fig8.eps}). The SHG is due to the presence of a weak quadratic nonlinearity because of the Ge doping ions.

\begin{figure}[htb!]
  \centering
    \includegraphics[scale=.7]{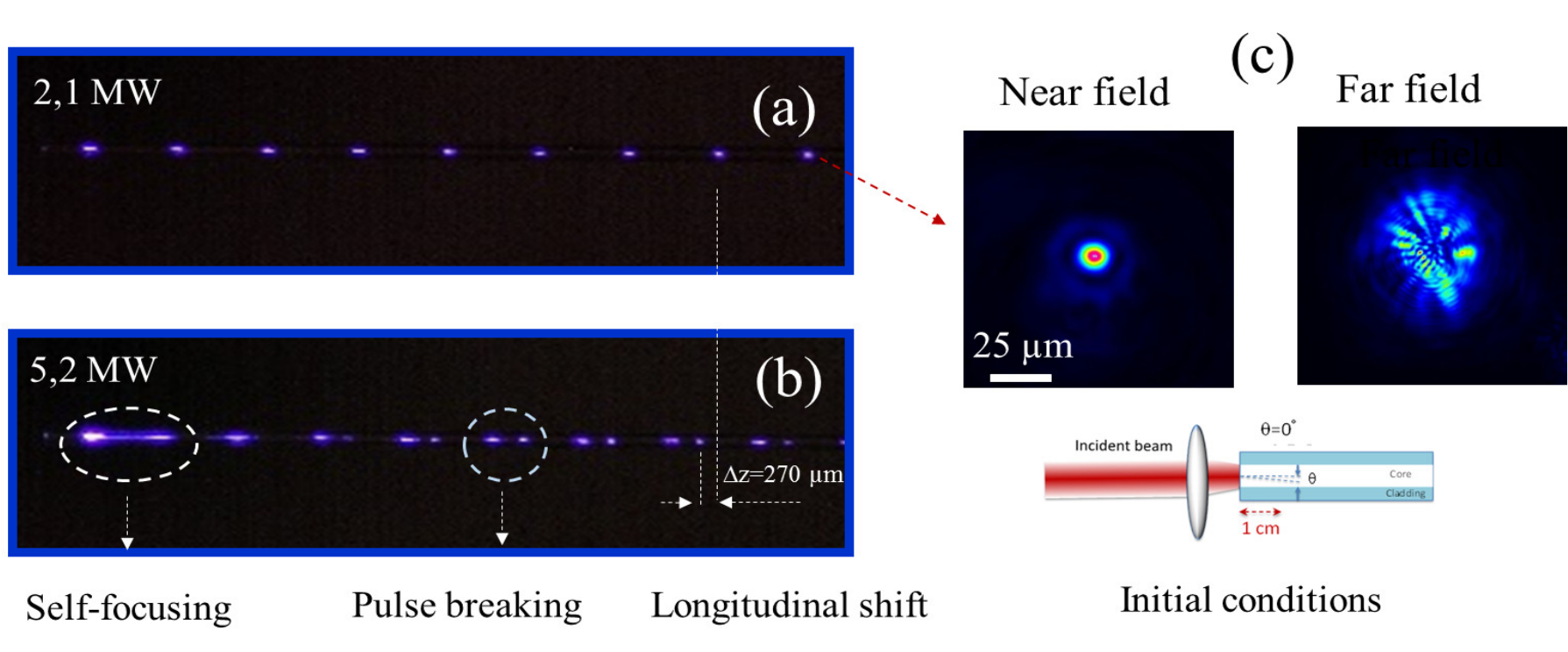}
  \caption{Experimental side images of a 50/125 optical fiber excited by a 250 fs laser source at 1030 nm showing the periodic self-imaging process; (a) image recorded at 2.1 MW; (b) 5.2 MW; (c) output near and far fields for an input peak power of 2.1 MW.}
    \label{fig9.eps}
    \end{figure}

    
Beyond the observation of the self-imaging periodicity by means of non-collinear frequency conversions, we investigated possible longitudinal distortions introduced by a self-focusing regime (see Figure~\ref{fig9.eps}). By increasing the input beam power (up to 5.2 MW), and keeping the on-axis excitation, we initiated a self-focusing propagation regime, starting from the first node of the self-imaging process. 

This spatial trapping, which resembles a Townes soliton \cite{PhysRevLett.13.479}, propagates over hundreds of micrometers before recovering its diffracting nature under the effect of nonlinear losses introduced by frequency conversions. Because of its intensity-dependent nature, this extreme event can drastically modulate the initial mode beating, by introducing both a pulse breaking process and a shift of the self-imaging periodicity. As a result, the intense beam of Figure~\ref{fig9.eps}(b) is transformed in a double peak of intensity, both oscillating with the same initial period of the self-imaging process. 
    
\begin{figure}[htb!]
  \centering
    \includegraphics[scale=.4]{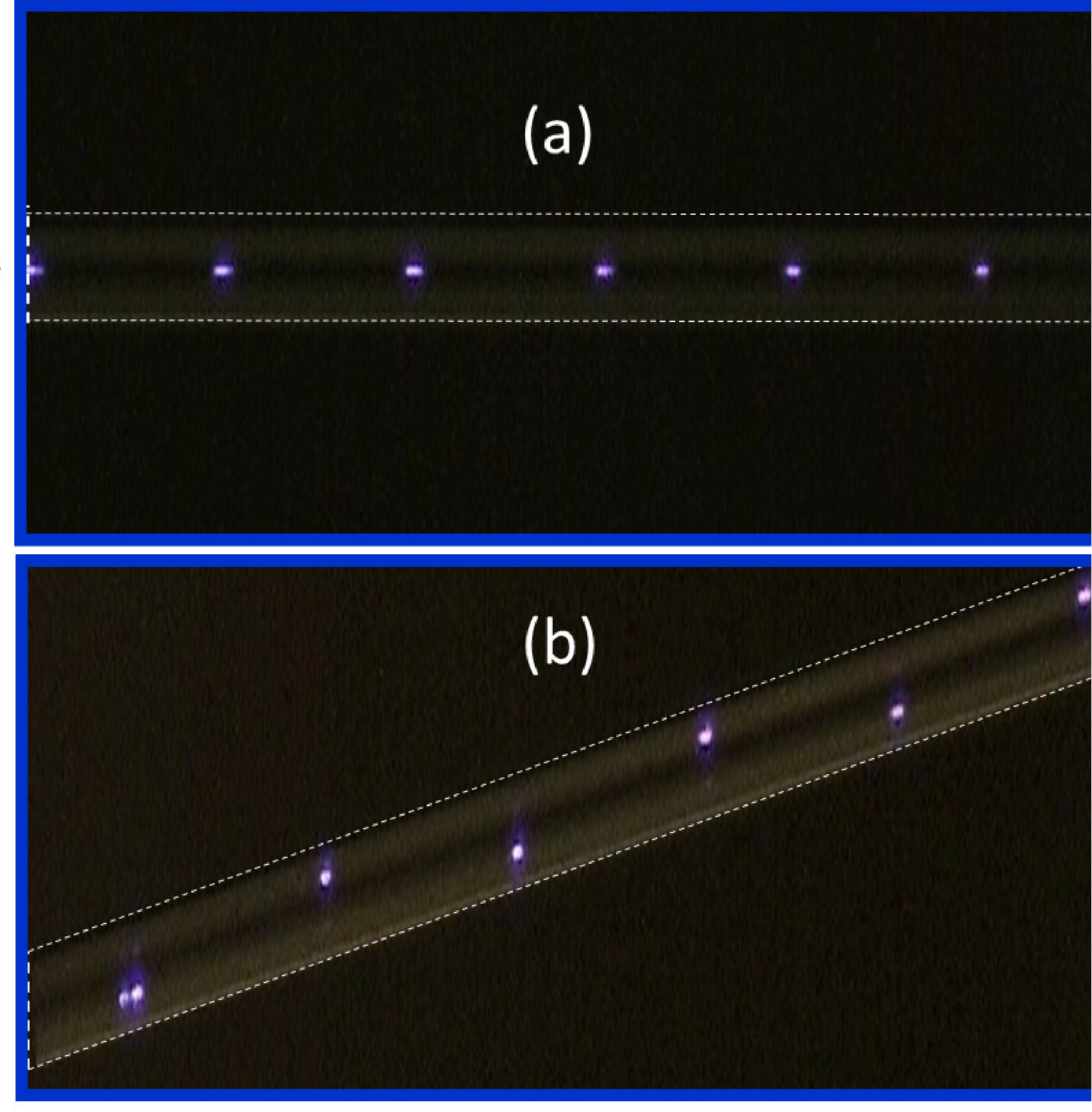}
  \caption{Experimental side images of a 50/125 optical fiber excited by a 250 fs laser source at 1030 nm for 2.1 MW of peak power; (a) axial coupling; (b) off-axis excitation.}
    \label{fig10.eps}
    \end{figure}

As shown by Figure~\ref{fig10.eps}, light scattering outside the fiber, at 1030 nm, gives again a clear visualization of the presence of SSI, which leads to sharp intensity peaks along the periodic evolution of the field in the GRIN MMF. The measured periodicity of the light intensity in the fiber matches well the value obtained with experiments at 1550 nm, as well as the theoretical value.

 We also investigated the case when the input beam is injected at a small angle with respect to the axis of the fiber. In this case, as shown by the resulting series of bright spots of Figure~\ref{fig10.eps}, the beam traces out a zig-zag trajectory, as it appears to be reflected by the core-cladding index boundary.

It is also clearly visible that the transverse dimension of the spot does not extend over all the transverse fiber core section, but is either localized on the central part or on a side. Thus, the longitudinal modulation imprinted on the propagating light inside the fiber is directly dependent of the initial coupling conditions, i.e. of the combination of initially excited modes which can self-modulate because of the mode beating and the Kerr effect, and initiate a spatial exchange of energy between them. This mechanism is at the origin of the spatial self-cleaning effect, which has been largely reported in the literature \cite{Krupanatphotonics,WrightNP2016}. Additionally, this experiment demonstrates that the transverse position of the nodes of the periodic beating can be transversely tuned by means of the initial coupling conditions. This brings an additional degree of freedom to realize a nonlinear saturable absorption mechanism, eventually favouring the emergence of a high-order mode, in order to benefit of the higher dispersion regime when light is coupled back in the multimode fiber. In this sense, temporal mode-locking on a high-order transverse mode should be possible.

 \section{Conclusion}
We studied the dynamics of beam-self imaging in nonlinear GRIN multimode optical fibers. We obtained an exact solution for the first and second order moments of a laser beam, describing both the period and the amplitude of the beam width oscillations along the fiber. The theory also permits to analytically predict the critical power for beam critical self-focusing, or collapse. We experimentally studied the longitudinal evolution of beam self-imaging by means of femtosecond laser pulse propagation in both the anomalous and the normal dispersion regime of a standard GRIN fiber. By observing light scattering out of the fiber core via visible fluorescence emission and harmonic wave generation, we could directly confirm that the invariance of the self-imaging period up to values close to beam collapse. These findings are of interest for applications involving fiber lasers mode-locked via multimode interference, and to all-optical beam processing with multimode fibers. \appendix

\section*{Funding}

European Research Council (ERC) (740355); Agence Nationale de la Recherche (ANR) (ANR-18-CE080016-01); CILAS Company (ArianeGroup) by means of the shared X-LAS laboratory; Swedish Research Council (Grant No. 2017-05309); Russian Ministry of Science and Education (14.Y26.31.0017); 

\section*{Acknowledgments}
We thank Fabrizio Frezza for helpful discussions.

\section*{Disclosures}


\noindent The authors declare no conflicts of interest.


\section{Appendix}




Momentum operators must satisfy certain commutation relations
\begin{equation}
  [\hat{x}_i,\hat{x}_j] = 0, \qquad [\hat{p}_i,\hat{p}_j] = 0, \qquad [\hat{x}_i,\hat{p}_j] = i\delta_{ij}, \qquad [\hat{x}_i,\hat{I}(x,y)] = 0,
\end{equation}
where $i,j$ denotes combinations of the $x,y$ operators, and the commutator is defined by $[\hat{f},\hat{g}] = \hat{f}\hat{g}-\hat{g}\hat{f}$. We assume that $\hbar = 1$ and that the commutators obey the standard Lie algebra
\begin{equation}
  [\hat{f},\hat{g}] = -[\hat{g},\hat{f}], \qquad [\hat{f},\alpha \hat{g}+\beta \hat{h}] = \alpha [\hat{f},\hat{g}]+\beta [\hat{f},\hat{h}], \qquad [\hat{f},[\hat{g},\hat{h}]]+[\hat{g},[\hat{h},\hat{f}]]+[\hat{h},[\hat{f},\hat{g}]] = 0,
\end{equation}
together with the relation $[\hat{f},\hat{g}\hat{h}] = \hat{g}[\hat{f},\hat{h}]+[\hat{f},\hat{g}]\hat{h}$.

To derive the moment equations we make use of the fact that the integrals can be expressed as expectation values of operators
\begin{equation}
  \langle q\rangle = \int A\hat{q}A^*~ds,
\end{equation}
where $ds = dx dy$. As a result, the evolution equation for each moment are obtained from
\begin{equation}
  \frac{d\langle q\rangle}{dz} = i\langle[\hat{q},\hat{H}]\rangle.
  \label{eq:evol}
\end{equation}
In particular we have that
\begin{equation}
  \langle x\rangle = \int x|A|^2~ds, \qquad \langle p_x\rangle = \frac{i}{2}\int\left(\frac{\partial A}{\partial x}A^*-A\frac{\partial A^*}{\partial x}\right)~ds, \qquad \langle I\rangle = \int |A|^4~ds,
\end{equation}
\begin{equation}
  \langle x^2\rangle = \int x^2|A|^2~ds, \qquad \langle xp_x+p_xx\rangle = i\int x\left(\frac{\partial A}{\partial x}A^*-A\frac{\partial A^*}{\partial x}\right)~ds, \qquad \langle p_x^2\rangle = \int \bigg|\frac{\partial A}{\partial x}\bigg|^2~ds,
\end{equation}
and analogous for $y$.

For the first-order moments we find that
\begin{align}
  & \frac{d\langle x\rangle}{dz} = -\frac{1}{k_0}\langle p_x\rangle, & \frac{d\langle p_x\rangle}{dz} = 2\beta\langle x\rangle, \\
  & \frac{d\langle y\rangle}{dz} = -\frac{1}{k_0}\langle p_y\rangle, & \frac{d\langle p_y\rangle}{dz} = 2\beta\langle y\rangle.
\end{align}
These equations have sinusoidal solutions (see Eqs.~(\ref{eq:sinus}))%

For the second-order moments we have that
\begin{align}
  & \frac{d\langle x^2\rangle}{dz} = -\frac{1}{k_0}\langle xp_x+p_xx\rangle, & \frac{d\langle y^2\rangle}{dz} = -\frac{1}{k_0}\langle yp_y+p_yy\rangle, \\
  & \frac{d\langle p_x^2\rangle}{dz} = 2\beta\langle xp_x+p_xx\rangle - i\gamma\langle [p_x^2,I]\rangle, & \frac{d\langle p_y^2\rangle}{dz} = 2\beta\langle yp_y+p_yy\rangle - i\gamma\langle [p_y^2,I]\rangle, \\
  & \frac{d\langle xp_x+p_xx\rangle}{dz} = -\frac{2}{k_0}\langle p_x^2\rangle + 4\beta\langle x^2\rangle + \gamma\langle I\rangle, & \frac{d\langle yp_y+p_yy\rangle}{dz} = -\frac{2}{k_0}\langle p_y^2\rangle + 4\beta\langle y^2\rangle + \gamma\langle I\rangle,
\end{align}
where we have used that $\langle[xp_x+p_xx,I]\rangle = \langle[yp_y+p_yy,I]\rangle = i\langle I\rangle$ and it remains to evaluate the commutators between the square momenta and $\hat{I}$. From the definition we have that
\begin{equation}
  \langle [p_x^2,I]\rangle = \int\frac{\partial (|A|^2)}{\partial x}\left(\frac{\partial A}{\partial x}A^*-A\frac{\partial A^*}{\partial x}\right)~ds,
\end{equation}
and similar for $y$. The commutators are related through (here some care is necessary due to the nonlinearity)
\begin{equation}
  \frac{d\langle I\rangle}{dz} = -i\frac{1}{k_0}\left(\langle [p_x^2,I]\rangle+\langle [p_y^2,I]\rangle\right).
\end{equation}

It can be verified that the proper Hamiltonian invariant
\begin{equation}
  \langle H_0\rangle =  \frac{1}{k_0}\left(\langle p_x^2\rangle+\langle p_y^2\rangle\right) + 2\beta\left(\langle x^2\rangle+\langle y^2\rangle\right)-\gamma\langle I\rangle
 \label{eq:H0}
  \end{equation}
is conserved. In addition to the Hamiltonian, we also have conservation of power and angular momentum, viz.
\begin{equation}
  \langle P_0\rangle = \int |A|^2~ds, \qquad \langle yp_x-xp_y\rangle = \frac{i}{2}\int\left[y\left(\frac{\partial A}{\partial x}A^*-A\frac{\partial A^*}{\partial x}\right)-x\left(\frac{\partial A}{\partial y}A^*-A\frac{\partial A^*}{\partial y}\right)\right]~ds.
\end{equation}

\subsection{Collapse conditions}

The integration constants of Eq.~(\ref{eq:rms}) can in general be determined from
\begin{equation}
  a_r^2 = \left(\langle x^2+y^2\rangle_0 - \frac{1}{4\beta}\langle H_0\rangle\right)^2 + \frac{k_0}{8\beta}\left(\frac{d\langle x^2+y^2\rangle_0}{dz}\right)^2,
\end{equation}
\begin{equation}
  \tan{b_r} = -\sqrt{\frac{k_0}{8\beta}}\frac{d\langle x^2+y^2\rangle_0}{dz}\bigg/\left(\langle x^2+y^2\rangle_0-\frac{1}{4\beta}\langle H_0\rangle\right),
\end{equation}
with the change in rms-width for the initial beam profile given by
\begin{equation}
  \frac{d\langle x^2+y^2\rangle_0}{dz} = -\frac{1}{k_0}\left[\langle xp_x+p_xx\rangle_0+\langle yp_y+p_yy\rangle_0\right].
\end{equation}

If the beam has a nonvanishing initial phase-front curvature, $b_r \neq 0$, we may rewrite Eq.~(\ref{eq:Crms}) as
\begin{equation}
  \langle x^2+y^2\rangle (z) = \frac{\langle x^2+y^2\rangle_0}{1+\cos{b_r}}\left[\left(1+C\cos{b_r}\right)+\left(1-C\right)\cos\left(\sqrt{\frac{8\beta}{k_0}}z+b_r\right)\right],
  \label{eq:Crms2}
\end{equation}
where the C-parameter satisfies
\begin{equation}
  C = \frac{1}{4\beta}\frac{\langle H_0\rangle}{\langle x^2+y^2\rangle_0} - \frac{1}{\langle x^2+y^2\rangle_0}\sqrt{\left(\langle x^2+y^2\rangle_0 - \frac{1}{4\beta}\langle H_0\rangle\right)^2 + \frac{k_0}{8\beta}\left(\frac{d\langle x^2+y^2\rangle_0}{dz}\right)^2},
\end{equation}
and is defined by the condition that the minimum rms-width is $\langle x^2+y^2\rangle = C\langle x^2+y^2\rangle_0$. Collapse will therefore occur at some point along the beam trajectory when $C \leq 0$.

\subsection{Supergaussian beam profile}

To investigate the influence of the beam shape on the dynamics we also consider a supergaussian beam profile

\begin{equation}
  A(x,y,0) = A_0\exp\left[-\frac{1}{2}\left(\frac{x}{x_0}\right)^4-\frac{1}{2}\left(\frac{y}{y_0}\right)^4\right].
\end{equation}
The power and Hamiltonian invariants for this case are obtain as
\begin{equation}
  \langle P_0\rangle = \frac{\Gamma^2(1/4)}{4}|A_0|^2x_0y_0, \qquad \langle H_0\rangle = \frac{4\pi\langle P_0\rangle}{\sqrt{2}\Gamma^2(1/4)}\left[\frac{3}{2k_0}\left(\frac{1}{x_0^2}+\frac{1}{y_0^2}\right)+\beta(x_0^2+y_0^2)-\frac{\gamma \langle P_0\rangle}{\pi x_0y_0}\right],
\end{equation}
while the beam width and C-parameter becomes
\begin{equation}
  w_f = \frac{2^{1/4}}{\Gamma(1/4)}\sqrt{\pi(x_0^2+y_0^2)}, \qquad C = \frac{3}{2\beta k_0x_0^2y_0^2}\left(1-\frac{2\gamma k_0\langle P_0\rangle x_0y_0}{3\pi(x_0^2+y_0^2)}\right).
\end{equation}
Also here it is seen that the critical collapse power is independent of the beam width for a symmetric beam.



\bibliographystyle{unsrt}

\bibliography{BIBLIO_MultimodeTemp3}

\begin{thebibliography}{27}%
\makeatletter
\providecommand \@ifxundefined [1]{%
 \@ifx{#1\undefined}
}%
\providecommand \@ifnum [1]{%
 \ifnum #1\expandafter \@firstoftwo
 \else \expandafter \@secondoftwo
 \fi
}%
\providecommand \@ifx [1]{%
 \ifx #1\expandafter \@firstoftwo
 \else \expandafter \@secondoftwo
 \fi
}%
\providecommand \natexlab [1]{#1}%
\providecommand \enquote  [1]{``#1''}%
\providecommand \bibnamefont  [1]{#1}%
\providecommand \bibfnamefont [1]{#1}%
\providecommand \citenamefont [1]{#1}%
\providecommand \href@noop [0]{\@secondoftwo}%
\providecommand \href [0]{\begingroup \@sanitize@url \@href}%
\providecommand \@href[1]{\@@startlink{#1}\@@href}%
\providecommand \@@href[1]{\endgroup#1\@@endlink}%
\providecommand \@sanitize@url [0]{\catcode `\\12\catcode `\$12\catcode
  `\&12\catcode `\#12\catcode `\^12\catcode `\_12\catcode `\%12\relax}%
\providecommand \@@startlink[1]{}%
\providecommand \@@endlink[0]{}%
\providecommand \url  [0]{\begingroup\@sanitize@url \@url }%
\providecommand \@url [1]{\endgroup\@href {#1}{\urlprefix }}%
\providecommand \urlprefix  [0]{URL }%
\providecommand \Eprint [0]{\href }%
\providecommand \doibase [0]{http://dx.doi.org/}%
\providecommand \selectlanguage [0]{\@gobble}%
\providecommand \bibinfo  [0]{\@secondoftwo}%
\providecommand \bibfield  [0]{\@secondoftwo}%
\providecommand \translation [1]{[#1]}%
\providecommand \BibitemOpen [0]{}%
\providecommand \bibitemStop [0]{}%
\providecommand \bibitemNoStop [0]{.\EOS\space}%
\providecommand \EOS [0]{\spacefactor3000\relax}%
\providecommand \BibitemShut  [1]{\csname bibitem#1\endcsname}%
\let\auto@bib@innerbib\@empty
\bibitem [{\citenamefont {Krupa}\ \emph {et~al.}(2019)\citenamefont {Krupa},
  \citenamefont {Tonello}, \citenamefont {Barth\'{e}l\'{e}my}, \citenamefont
  {Mansuryan}, \citenamefont {Couderc}, \citenamefont {Millot}, \citenamefont
  {Grelu}, \citenamefont {Modotto}, \citenamefont {Babin},\ and\ \citenamefont
  {Wabnitz}}]{doi:10.1063/1.5119434}%
  \BibitemOpen
  \bibfield  {author} {\bibinfo {author} {\bibfnamefont {K.}~\bibnamefont
  {Krupa}}, \bibinfo {author} {\bibfnamefont {A.}~\bibnamefont {Tonello}},
  \bibinfo {author} {\bibfnamefont {A.}~\bibnamefont {Barth\'{e}l\'{e}my}},
  \bibinfo {author} {\bibfnamefont {T.}~\bibnamefont {Mansuryan}}, \bibinfo
  {author} {\bibfnamefont {V.}~\bibnamefont {Couderc}}, \bibinfo {author}
  {\bibfnamefont {G.}~\bibnamefont {Millot}}, \bibinfo {author} {\bibfnamefont
  {P.}~\bibnamefont {Grelu}}, \bibinfo {author} {\bibfnamefont
  {D.}~\bibnamefont {Modotto}}, \bibinfo {author} {\bibfnamefont {S.~A.}\
  \bibnamefont {Babin}}, \ and\ \bibinfo {author} {\bibfnamefont
  {S.}~\bibnamefont {Wabnitz}},\ }\href {\doibase 10.1063/1.5119434} {\bibfield
   {journal} {\bibinfo  {journal} {APL Photonics}\ }\textbf {\bibinfo {volume}
  {4}},\ \bibinfo {pages} {110901} (\bibinfo {year} {2019})},\ \Eprint
  {http://arxiv.org/abs/https://doi.org/10.1063/1.5119434}
  {https://doi.org/10.1063/1.5119434} \BibitemShut {NoStop}%
\bibitem [{\citenamefont {Picozzi}\ \emph {et~al.}(2015)\citenamefont
  {Picozzi}, \citenamefont {Millot},\ and\ \citenamefont
  {Wabnitz}}]{Picozzi2015R30}%
  \BibitemOpen
  \bibfield  {author} {\bibinfo {author} {\bibfnamefont {A.}~\bibnamefont
  {Picozzi}}, \bibinfo {author} {\bibfnamefont {G.}~\bibnamefont {Millot}}, \
  and\ \bibinfo {author} {\bibfnamefont {S.}~\bibnamefont {Wabnitz}},\
  }\href@noop {} {\bibfield  {journal} {\bibinfo  {journal} {Nat. Photonics}\
  }\textbf {\bibinfo {volume} {9}},\ \bibinfo {pages} {289} (\bibinfo {year}
  {2015})}\BibitemShut {NoStop}%
\bibitem [{\citenamefont {{Mafi}}(2012)}]{6242367}%
  \BibitemOpen
  \bibfield  {author} {\bibinfo {author} {\bibfnamefont {A.}~\bibnamefont
  {{Mafi}}},\ }\href {\doibase 10.1109/JLT.2012.2208215} {\bibfield  {journal}
  {\bibinfo  {journal} {Journal of Lightwave Technology}\ }\textbf {\bibinfo
  {volume} {30}},\ \bibinfo {pages} {2803} (\bibinfo {year}
  {2012})}\BibitemShut {NoStop}%
\bibitem [{\citenamefont {Talbot}(1836)}]{doi:10.1080/14786443608649032}%
  \BibitemOpen
  \bibfield  {author} {\bibinfo {author} {\bibfnamefont {H.}~\bibnamefont
  {Talbot}},\ }\href@noop {} {\bibfield  {journal} {\bibinfo  {journal} {The
  London, Edinburgh, and Dublin Philosophical Magazine and Journal of Science}\
  }\textbf {\bibinfo {volume} {9}},\ \bibinfo {pages} {401} (\bibinfo {year}
  {1836})}\BibitemShut {NoStop}%
\bibitem [{\citenamefont {Agrawal}(2019)}]{AGRAWAL2019309}%
  \BibitemOpen
  \bibfield  {author} {\bibinfo {author} {\bibfnamefont {G.~P.}\ \bibnamefont
  {Agrawal}},\ }\href {\doibase https://doi.org/10.1016/j.yofte.2019.04.012}
  {\bibfield  {journal} {\bibinfo  {journal} {Optical Fiber Technology}\
  }\textbf {\bibinfo {volume} {50}},\ \bibinfo {pages} {309 } (\bibinfo {year}
  {2019})}\BibitemShut {NoStop}%
\bibitem [{\citenamefont {Wright}\ \emph
  {et~al.}(2015{\natexlab{a}})\citenamefont {Wright}, \citenamefont
  {Christodoulides},\ and\ \citenamefont {Wise}}]{Wright2015R31}%
  \BibitemOpen
  \bibfield  {author} {\bibinfo {author} {\bibfnamefont {L.~G.}\ \bibnamefont
  {Wright}}, \bibinfo {author} {\bibfnamefont {D.~N.}\ \bibnamefont
  {Christodoulides}}, \ and\ \bibinfo {author} {\bibfnamefont {F.~W.}\
  \bibnamefont {Wise}},\ }\href@noop {} {\bibfield  {journal} {\bibinfo
  {journal} {Nat. Photonics}\ }\textbf {\bibinfo {volume} {9}},\ \bibinfo
  {pages} {306} (\bibinfo {year} {2015}{\natexlab{a}})}\BibitemShut {NoStop}%
\bibitem [{\citenamefont {Wright}\ \emph
  {et~al.}(2015{\natexlab{b}})\citenamefont {Wright}, \citenamefont {Wabnitz},
  \citenamefont {Christodoulides},\ and\ \citenamefont
  {Wise}}]{PhysRevLett.115.223902}%
  \BibitemOpen
  \bibfield  {author} {\bibinfo {author} {\bibfnamefont {L.~G.}\ \bibnamefont
  {Wright}}, \bibinfo {author} {\bibfnamefont {S.}~\bibnamefont {Wabnitz}},
  \bibinfo {author} {\bibfnamefont {D.~N.}\ \bibnamefont {Christodoulides}}, \
  and\ \bibinfo {author} {\bibfnamefont {F.~W.}\ \bibnamefont {Wise}},\ }\href
  {\doibase 10.1103/PhysRevLett.115.223902} {\bibfield  {journal} {\bibinfo
  {journal} {Phys. Rev. Lett.}\ }\textbf {\bibinfo {volume} {115}},\ \bibinfo
  {pages} {223902} (\bibinfo {year} {2015}{\natexlab{b}})}\BibitemShut
  {NoStop}%
\bibitem [{\citenamefont {Krupa}\ \emph {et~al.}(2016)\citenamefont {Krupa},
  \citenamefont {Tonello}, \citenamefont {Barth\'el\'emy}, \citenamefont
  {Couderc}, \citenamefont {Shalaby}, \citenamefont {Bendahmane}, \citenamefont
  {Millot},\ and\ \citenamefont {Wabnitz}}]{PhysRevLett.116.183901}%
  \BibitemOpen
  \bibfield  {author} {\bibinfo {author} {\bibfnamefont {K.}~\bibnamefont
  {Krupa}}, \bibinfo {author} {\bibfnamefont {A.}~\bibnamefont {Tonello}},
  \bibinfo {author} {\bibfnamefont {A.}~\bibnamefont {Barth\'el\'emy}},
  \bibinfo {author} {\bibfnamefont {V.}~\bibnamefont {Couderc}}, \bibinfo
  {author} {\bibfnamefont {B.~M.}\ \bibnamefont {Shalaby}}, \bibinfo {author}
  {\bibfnamefont {A.}~\bibnamefont {Bendahmane}}, \bibinfo {author}
  {\bibfnamefont {G.}~\bibnamefont {Millot}}, \ and\ \bibinfo {author}
  {\bibfnamefont {S.}~\bibnamefont {Wabnitz}},\ }\href {\doibase
  10.1103/PhysRevLett.116.183901} {\bibfield  {journal} {\bibinfo  {journal}
  {Phys. Rev. Lett.}\ }\textbf {\bibinfo {volume} {116}},\ \bibinfo {pages}
  {183901} (\bibinfo {year} {2016})}\BibitemShut {NoStop}%
\bibitem [{\citenamefont {Krupa}\ \emph {et~al.}(2017)\citenamefont {Krupa},
  \citenamefont {Tonello}, \citenamefont {Shalaby}, \citenamefont {Fabert},
  \citenamefont {Barth{\'e}l{\'e}my}, \citenamefont {Millot}, \citenamefont
  {Wabnitz},\ and\ \citenamefont {Couderc}}]{Krupanatphotonics}%
  \BibitemOpen
  \bibfield  {author} {\bibinfo {author} {\bibfnamefont {K.}~\bibnamefont
  {Krupa}}, \bibinfo {author} {\bibfnamefont {A.}~\bibnamefont {Tonello}},
  \bibinfo {author} {\bibfnamefont {B.~M.}\ \bibnamefont {Shalaby}}, \bibinfo
  {author} {\bibfnamefont {M.}~\bibnamefont {Fabert}}, \bibinfo {author}
  {\bibfnamefont {A.}~\bibnamefont {Barth{\'e}l{\'e}my}}, \bibinfo {author}
  {\bibfnamefont {G.}~\bibnamefont {Millot}}, \bibinfo {author} {\bibfnamefont
  {S.}~\bibnamefont {Wabnitz}}, \ and\ \bibinfo {author} {\bibfnamefont
  {V.}~\bibnamefont {Couderc}},\ }\href@noop {} {\bibfield  {journal} {\bibinfo
   {journal} {Nat. Photonics}\ }\textbf {\bibinfo {volume} {11}},\ \bibinfo
  {pages} {234} (\bibinfo {year} {2017})}\BibitemShut {NoStop}%
\bibitem [{\citenamefont {Wright}\ \emph {et~al.}(2016)\citenamefont {Wright},
  \citenamefont {Liu}, \citenamefont {Nolan}, \citenamefont {Li}, \citenamefont
  {Christodoulides},\ and\ \citenamefont {Wise}}]{WrightNP2016}%
  \BibitemOpen
  \bibfield  {author} {\bibinfo {author} {\bibfnamefont {L.~G.}\ \bibnamefont
  {Wright}}, \bibinfo {author} {\bibfnamefont {Z.}~\bibnamefont {Liu}},
  \bibinfo {author} {\bibfnamefont {D.~A.}\ \bibnamefont {Nolan}}, \bibinfo
  {author} {\bibfnamefont {M.-J.}\ \bibnamefont {Li}}, \bibinfo {author}
  {\bibfnamefont {D.~N.}\ \bibnamefont {Christodoulides}}, \ and\ \bibinfo
  {author} {\bibfnamefont {F.~W.}\ \bibnamefont {Wise}},\ }\href@noop {}
  {\bibfield  {journal} {\bibinfo  {journal} {Nat. Photonics}\ }\textbf
  {\bibinfo {volume} {10}},\ \bibinfo {pages} {771} (\bibinfo {year}
  {2016})}\BibitemShut {NoStop}%
\bibitem [{\citenamefont {Zhu}\ \emph {et~al.}(2008)\citenamefont {Zhu},
  \citenamefont {Sch\"{u}lzgen}, \citenamefont {Li}, \citenamefont {Li},
  \citenamefont {Han}, \citenamefont {Moloney},\ and\ \citenamefont
  {Peyghambarian}}]{Zhu:08}%
  \BibitemOpen
  \bibfield  {author} {\bibinfo {author} {\bibfnamefont {X.}~\bibnamefont
  {Zhu}}, \bibinfo {author} {\bibfnamefont {A.}~\bibnamefont {Sch\"{u}lzgen}},
  \bibinfo {author} {\bibfnamefont {H.}~\bibnamefont {Li}}, \bibinfo {author}
  {\bibfnamefont {L.}~\bibnamefont {Li}}, \bibinfo {author} {\bibfnamefont
  {L.}~\bibnamefont {Han}}, \bibinfo {author} {\bibfnamefont {J.~V.}\
  \bibnamefont {Moloney}}, \ and\ \bibinfo {author} {\bibfnamefont
  {N.}~\bibnamefont {Peyghambarian}},\ }\href {\doibase 10.1364/OE.16.016632}
  {\bibfield  {journal} {\bibinfo  {journal} {Opt. Express}\ }\textbf {\bibinfo
  {volume} {16}},\ \bibinfo {pages} {16632} (\bibinfo {year}
  {2008})}\BibitemShut {NoStop}%
\bibitem [{\citenamefont {Nazemosadat}\ and\ \citenamefont
  {Mafi}(2013)}]{Nazemosadat:13}%
  \BibitemOpen
  \bibfield  {author} {\bibinfo {author} {\bibfnamefont {E.}~\bibnamefont
  {Nazemosadat}}\ and\ \bibinfo {author} {\bibfnamefont {A.}~\bibnamefont
  {Mafi}},\ }\href {\doibase 10.1364/JOSAB.30.001357} {\bibfield  {journal}
  {\bibinfo  {journal} {J. Opt. Soc. Am. B}\ }\textbf {\bibinfo {volume}
  {30}},\ \bibinfo {pages} {1357} (\bibinfo {year} {2013})}\BibitemShut
  {NoStop}%
\bibitem [{\citenamefont {Poletti}\ and\ \citenamefont
  {Horak}(2008)}]{Poletti:08}%
  \BibitemOpen
  \bibfield  {author} {\bibinfo {author} {\bibfnamefont {F.}~\bibnamefont
  {Poletti}}\ and\ \bibinfo {author} {\bibfnamefont {P.}~\bibnamefont
  {Horak}},\ }\href {\doibase 10.1364/JOSAB.25.001645} {\bibfield  {journal}
  {\bibinfo  {journal} {J. Opt. Soc. Am. B}\ }\textbf {\bibinfo {volume}
  {25}},\ \bibinfo {pages} {1645} (\bibinfo {year} {2008})}\BibitemShut
  {NoStop}%
\bibitem [{\citenamefont {Fu}\ \emph {et~al.}(2016)\citenamefont {Fu},
  \citenamefont {Shi}, \citenamefont {Sheng}, \citenamefont {Shi},
  \citenamefont {Zhu}, \citenamefont {Yao}, \citenamefont {Norwood},\ and\
  \citenamefont {Peyghambarian}}]{Fu:16}%
  \BibitemOpen
  \bibfield  {author} {\bibinfo {author} {\bibfnamefont {S.}~\bibnamefont
  {Fu}}, \bibinfo {author} {\bibfnamefont {G.}~\bibnamefont {Shi}}, \bibinfo
  {author} {\bibfnamefont {Q.}~\bibnamefont {Sheng}}, \bibinfo {author}
  {\bibfnamefont {W.}~\bibnamefont {Shi}}, \bibinfo {author} {\bibfnamefont
  {X.}~\bibnamefont {Zhu}}, \bibinfo {author} {\bibfnamefont {J.}~\bibnamefont
  {Yao}}, \bibinfo {author} {\bibfnamefont {R.~A.}\ \bibnamefont {Norwood}}, \
  and\ \bibinfo {author} {\bibfnamefont {N.}~\bibnamefont {Peyghambarian}},\
  }\href {\doibase 10.1364/OE.24.011282} {\bibfield  {journal} {\bibinfo
  {journal} {Opt. Express}\ }\textbf {\bibinfo {volume} {24}},\ \bibinfo
  {pages} {11282} (\bibinfo {year} {2016})}\BibitemShut {NoStop}%
\bibitem [{\citenamefont {{Wang}}\ \emph {et~al.}(2017)\citenamefont {{Wang}},
  \citenamefont {{Wang}}, \citenamefont {{Yang}}, \citenamefont {{Li}},
  \citenamefont {{Zhao}}, \citenamefont {{Xu}}, \citenamefont {{Jin}},
  \citenamefont {{Cao}},\ and\ \citenamefont {{Fang}}}]{8093610}%
  \BibitemOpen
  \bibfield  {author} {\bibinfo {author} {\bibfnamefont {Z.}~\bibnamefont
  {{Wang}}}, \bibinfo {author} {\bibfnamefont {D.~N.}\ \bibnamefont {{Wang}}},
  \bibinfo {author} {\bibfnamefont {F.}~\bibnamefont {{Yang}}}, \bibinfo
  {author} {\bibfnamefont {L.}~\bibnamefont {{Li}}}, \bibinfo {author}
  {\bibfnamefont {C.}~\bibnamefont {{Zhao}}}, \bibinfo {author} {\bibfnamefont
  {B.}~\bibnamefont {{Xu}}}, \bibinfo {author} {\bibfnamefont {S.}~\bibnamefont
  {{Jin}}}, \bibinfo {author} {\bibfnamefont {S.}~\bibnamefont {{Cao}}}, \ and\
  \bibinfo {author} {\bibfnamefont {Z.}~\bibnamefont {{Fang}}},\ }\href
  {\doibase 10.1109/JLT.2017.2768663} {\bibfield  {journal} {\bibinfo
  {journal} {Journal of Lightwave Technology}\ }\textbf {\bibinfo {volume}
  {35}},\ \bibinfo {pages} {5280} (\bibinfo {year} {2017})}\BibitemShut
  {NoStop}%
\bibitem [{\citenamefont {Li}\ \emph {et~al.}(2017)\citenamefont {Li},
  \citenamefont {Wang}, \citenamefont {Li}, \citenamefont {Zhang},\ and\
  \citenamefont {Xu}}]{Li:17}%
  \BibitemOpen
  \bibfield  {author} {\bibinfo {author} {\bibfnamefont {H.}~\bibnamefont
  {Li}}, \bibinfo {author} {\bibfnamefont {Z.}~\bibnamefont {Wang}}, \bibinfo
  {author} {\bibfnamefont {C.}~\bibnamefont {Li}}, \bibinfo {author}
  {\bibfnamefont {J.}~\bibnamefont {Zhang}}, \ and\ \bibinfo {author}
  {\bibfnamefont {S.}~\bibnamefont {Xu}},\ }\href {\doibase
  10.1364/OE.25.026546} {\bibfield  {journal} {\bibinfo  {journal} {Opt.
  Express}\ }\textbf {\bibinfo {volume} {25}},\ \bibinfo {pages} {26546}
  (\bibinfo {year} {2017})}\BibitemShut {NoStop}%
\bibitem [{\citenamefont {Yang}\ \emph {et~al.}(2018)\citenamefont {Yang},
  \citenamefont {Wang}, \citenamefont {Wang}, \citenamefont {Li}, \citenamefont
  {Zhao}, \citenamefont {Xu}, \citenamefont {Jin}, \citenamefont {Cao},\ and\
  \citenamefont {Fang}}]{Yang:18}%
  \BibitemOpen
  \bibfield  {author} {\bibinfo {author} {\bibfnamefont {F.}~\bibnamefont
  {Yang}}, \bibinfo {author} {\bibfnamefont {D.~N.}\ \bibnamefont {Wang}},
  \bibinfo {author} {\bibfnamefont {Z.}~\bibnamefont {Wang}}, \bibinfo {author}
  {\bibfnamefont {L.}~\bibnamefont {Li}}, \bibinfo {author} {\bibfnamefont
  {C.-L.}\ \bibnamefont {Zhao}}, \bibinfo {author} {\bibfnamefont
  {B.}~\bibnamefont {Xu}}, \bibinfo {author} {\bibfnamefont {S.}~\bibnamefont
  {Jin}}, \bibinfo {author} {\bibfnamefont {S.-Y.}\ \bibnamefont {Cao}}, \ and\
  \bibinfo {author} {\bibfnamefont {Z.-J.}\ \bibnamefont {Fang}},\ }\href
  {\doibase 10.1364/OE.26.000927} {\bibfield  {journal} {\bibinfo  {journal}
  {Opt. Express}\ }\textbf {\bibinfo {volume} {26}},\ \bibinfo {pages} {927}
  (\bibinfo {year} {2018})}\BibitemShut {NoStop}%
\bibitem [{\citenamefont {Wang}\ \emph {et~al.}(2018)\citenamefont {Wang},
  \citenamefont {Wang}, \citenamefont {Yang}, \citenamefont {Li}, \citenamefont
  {Zhao}, \citenamefont {Xu}, \citenamefont {Jin}, \citenamefont {Cao},\ and\
  \citenamefont {Fang}}]{Wang:18a}%
  \BibitemOpen
  \bibfield  {author} {\bibinfo {author} {\bibfnamefont {Z.}~\bibnamefont
  {Wang}}, \bibinfo {author} {\bibfnamefont {D.~N.}\ \bibnamefont {Wang}},
  \bibinfo {author} {\bibfnamefont {F.}~\bibnamefont {Yang}}, \bibinfo {author}
  {\bibfnamefont {L.}~\bibnamefont {Li}}, \bibinfo {author} {\bibfnamefont
  {C.-L.}\ \bibnamefont {Zhao}}, \bibinfo {author} {\bibfnamefont
  {B.}~\bibnamefont {Xu}}, \bibinfo {author} {\bibfnamefont {S.}~\bibnamefont
  {Jin}}, \bibinfo {author} {\bibfnamefont {S.-Y.}\ \bibnamefont {Cao}}, \ and\
  \bibinfo {author} {\bibfnamefont {Z.-J.}\ \bibnamefont {Fang}},\ }\href
  {\doibase 10.1364/OL.43.002078} {\bibfield  {journal} {\bibinfo  {journal}
  {Opt. Lett.}\ }\textbf {\bibinfo {volume} {43}},\ \bibinfo {pages} {2078}
  (\bibinfo {year} {2018})}\BibitemShut {NoStop}%
\bibitem [{\citenamefont {Te\u{g}in}\ and\ \citenamefont
  {Orta\c{c}}(2018)}]{Tegin:18}%
  \BibitemOpen
  \bibfield  {author} {\bibinfo {author} {\bibfnamefont {U.}~\bibnamefont
  {Te\u{g}in}}\ and\ \bibinfo {author} {\bibfnamefont {B.}~\bibnamefont
  {Orta\c{c}}},\ }\href {\doibase 10.1364/OL.43.001611} {\bibfield  {journal}
  {\bibinfo  {journal} {Opt. Lett.}\ }\textbf {\bibinfo {volume} {43}},\
  \bibinfo {pages} {1611} (\bibinfo {year} {2018})}\BibitemShut {NoStop}%
\bibitem [{\citenamefont {Zhao}\ \emph
  {et~al.}(2018{\natexlab{a}})\citenamefont {Zhao}, \citenamefont {Wang},
  \citenamefont {Wang}, \citenamefont {Hu}, \citenamefont {Zhang},
  \citenamefont {Zhang},\ and\ \citenamefont {Cai}}]{Zhao_2018}%
  \BibitemOpen
  \bibfield  {author} {\bibinfo {author} {\bibfnamefont {F.}~\bibnamefont
  {Zhao}}, \bibinfo {author} {\bibfnamefont {Y.}~\bibnamefont {Wang}}, \bibinfo
  {author} {\bibfnamefont {H.}~\bibnamefont {Wang}}, \bibinfo {author}
  {\bibfnamefont {X.}~\bibnamefont {Hu}}, \bibinfo {author} {\bibfnamefont
  {W.}~\bibnamefont {Zhang}}, \bibinfo {author} {\bibfnamefont
  {T.}~\bibnamefont {Zhang}}, \ and\ \bibinfo {author} {\bibfnamefont
  {Y.}~\bibnamefont {Cai}},\ }\href {\doibase 10.1088/1555-6611/aac538}
  {\bibfield  {journal} {\bibinfo  {journal} {Laser Physics}\ }\textbf
  {\bibinfo {volume} {28}},\ \bibinfo {pages} {085104} (\bibinfo {year}
  {2018}{\natexlab{a}})}\BibitemShut {NoStop}%
\bibitem [{\citenamefont {Zhao}\ \emph
  {et~al.}(2018{\natexlab{b}})\citenamefont {Zhao}, \citenamefont {Wang},
  \citenamefont {Wang}, \citenamefont {Yan}, \citenamefont {Hu}, \citenamefont
  {Zhang}, \citenamefont {Zhang},\ and\ \citenamefont {Zhou}}]{Zhao:2018}%
  \BibitemOpen
  \bibfield  {author} {\bibinfo {author} {\bibfnamefont {F.}~\bibnamefont
  {Zhao}}, \bibinfo {author} {\bibfnamefont {Y.}~\bibnamefont {Wang}}, \bibinfo
  {author} {\bibfnamefont {H.}~\bibnamefont {Wang}}, \bibinfo {author}
  {\bibfnamefont {Z.}~\bibnamefont {Yan}}, \bibinfo {author} {\bibfnamefont
  {X.}~\bibnamefont {Hu}}, \bibinfo {author} {\bibfnamefont {W.}~\bibnamefont
  {Zhang}}, \bibinfo {author} {\bibfnamefont {T.}~\bibnamefont {Zhang}}, \ and\
  \bibinfo {author} {\bibfnamefont {K.}~\bibnamefont {Zhou}},\ }\href@noop {}
  {\bibfield  {journal} {\bibinfo  {journal} {Scientific Reports.}\ }\textbf
  {\bibinfo {volume} {8}},\ \bibinfo {pages} {16369} (\bibinfo {year}
  {2018}{\natexlab{b}})}\BibitemShut {NoStop}%
\bibitem [{\citenamefont {Chen}\ \emph {et~al.}(2019)\citenamefont {Chen},
  \citenamefont {Li}, \citenamefont {Wang}, \citenamefont {Zhang},
  \citenamefont {Zeng},\ and\ \citenamefont {Zhao}}]{Chen:19}%
  \BibitemOpen
  \bibfield  {author} {\bibinfo {author} {\bibfnamefont {G.}~\bibnamefont
  {Chen}}, \bibinfo {author} {\bibfnamefont {W.}~\bibnamefont {Li}}, \bibinfo
  {author} {\bibfnamefont {G.}~\bibnamefont {Wang}}, \bibinfo {author}
  {\bibfnamefont {W.}~\bibnamefont {Zhang}}, \bibinfo {author} {\bibfnamefont
  {C.}~\bibnamefont {Zeng}}, \ and\ \bibinfo {author} {\bibfnamefont
  {W.}~\bibnamefont {Zhao}},\ }\href {\doibase 10.1364/PRJ.7.000187} {\bibfield
   {journal} {\bibinfo  {journal} {Photon. Res.}\ }\textbf {\bibinfo {volume}
  {7}},\ \bibinfo {pages} {187} (\bibinfo {year} {2019})}\BibitemShut {NoStop}%
\bibitem [{\citenamefont {Karlsson}\ \emph {et~al.}(1992)\citenamefont
  {Karlsson}, \citenamefont {Anderson},\ and\ \citenamefont
  {Desaix}}]{Karlsson:92}%
  \BibitemOpen
  \bibfield  {author} {\bibinfo {author} {\bibfnamefont {M.}~\bibnamefont
  {Karlsson}}, \bibinfo {author} {\bibfnamefont {D.}~\bibnamefont {Anderson}},
  \ and\ \bibinfo {author} {\bibfnamefont {M.}~\bibnamefont {Desaix}},\ }\href
  {\doibase 10.1364/OL.17.000022} {\bibfield  {journal} {\bibinfo  {journal}
  {Opt. Lett.}\ }\textbf {\bibinfo {volume} {17}},\ \bibinfo {pages} {22}
  (\bibinfo {year} {1992})}\BibitemShut {NoStop}%
\bibitem [{\citenamefont {Cerenkov}(1934)}]{Cherenkov:1934}%
  \BibitemOpen
  \bibfield  {author} {\bibinfo {author} {\bibfnamefont {P.}~\bibnamefont
  {Cerenkov}},\ }\href@noop {} {\bibfield  {journal} {\bibinfo  {journal}
  {Dokl, Akad, Nauk, SSSR}\ }\textbf {\bibinfo {volume} {2}},\ \bibinfo {pages}
  {451} (\bibinfo {year} {1934})}\BibitemShut {NoStop}%
\bibitem [{\citenamefont {Dragoman}(1996)}]{Dragoman:96}%
  \BibitemOpen
  \bibfield  {author} {\bibinfo {author} {\bibfnamefont {D.}~\bibnamefont
  {Dragoman}},\ }\href {\doibase 10.1364/AO.35.004142} {\bibfield  {journal}
  {\bibinfo  {journal} {Appl. Opt.}\ }\textbf {\bibinfo {volume} {35}},\
  \bibinfo {pages} {4142} (\bibinfo {year} {1996})}\BibitemShut {NoStop}%
\bibitem [{\citenamefont {Hansson}\ \emph {et~al.}(2011)\citenamefont
  {Hansson}, \citenamefont {Anderson}, \citenamefont {Desaix},\ and\
  \citenamefont {Lisak}}]{HANSSON20113422}%
  \BibitemOpen
  \bibfield  {author} {\bibinfo {author} {\bibfnamefont {T.}~\bibnamefont
  {Hansson}}, \bibinfo {author} {\bibfnamefont {D.}~\bibnamefont {Anderson}},
  \bibinfo {author} {\bibfnamefont {M.}~\bibnamefont {Desaix}}, \ and\ \bibinfo
  {author} {\bibfnamefont {M.}~\bibnamefont {Lisak}},\ }\href {\doibase
  https://doi.org/10.1016/j.optcom.2011.03.013} {\bibfield  {journal} {\bibinfo
   {journal} {Optics Communications}\ }\textbf {\bibinfo {volume} {284}},\
  \bibinfo {pages} {3422 } (\bibinfo {year} {2011})}\BibitemShut {NoStop}%
\bibitem [{\citenamefont {Chiao}\ \emph {et~al.}(1964)\citenamefont {Chiao},
  \citenamefont {Garmire},\ and\ \citenamefont {Townes}}]{PhysRevLett.13.479}%
  \BibitemOpen
  \bibfield  {author} {\bibinfo {author} {\bibfnamefont {R.~Y.}\ \bibnamefont
  {Chiao}}, \bibinfo {author} {\bibfnamefont {E.}~\bibnamefont {Garmire}}, \
  and\ \bibinfo {author} {\bibfnamefont {C.~H.}\ \bibnamefont {Townes}},\
  }\href {\doibase 10.1103/PhysRevLett.13.479} {\bibfield  {journal} {\bibinfo
  {journal} {Phys. Rev. Lett.}\ }\textbf {\bibinfo {volume} {13}},\ \bibinfo
  {pages} {479} (\bibinfo {year} {1964})}\BibitemShut {NoStop}%
\end{thebibliography}%


\begin{thebibliography}{0}%
\makeatletter
\providecommand \@ifxundefined [1]{%
 \@ifx{#1\undefined}
}%
\providecommand \@ifnum [1]{%
 \ifnum #1\expandafter \@firstoftwo
 \else \expandafter \@secondoftwo
 \fi
}%
\providecommand \@ifx [1]{%
 \ifx #1\expandafter \@firstoftwo
 \else \expandafter \@secondoftwo
 \fi
}%
\providecommand \natexlab [1]{#1}%
\providecommand \enquote  [1]{``#1''}%
\providecommand \bibnamefont  [1]{#1}%
\providecommand \bibfnamefont [1]{#1}%
\providecommand \citenamefont [1]{#1}%
\providecommand \href@noop [0]{\@secondoftwo}%
\providecommand \href [0]{\begingroup \@sanitize@url \@href}%
\providecommand \@href[1]{\@@startlink{#1}\@@href}%
\providecommand \@@href[1]{\endgroup#1\@@endlink}%
\providecommand \@sanitize@url [0]{\catcode `\\12\catcode `\$12\catcode
  `\&12\catcode `\#12\catcode `\^12\catcode `\_12\catcode `\%12\relax}%
\providecommand \@@startlink[1]{}%
\providecommand \@@endlink[0]{}%
\providecommand \url  [0]{\begingroup\@sanitize@url \@url }%
\providecommand \@url [1]{\endgroup\@href {#1}{\urlprefix }}%
\providecommand \urlprefix  [0]{URL }%
\providecommand \Eprint [0]{\href }%
\providecommand \doibase [0]{http://dx.doi.org/}%
\providecommand \selectlanguage [0]{\@gobble}%
\providecommand \bibinfo  [0]{\@secondoftwo}%
\providecommand \bibfield  [0]{\@secondoftwo}%
\providecommand \translation [1]{[#1]}%
\providecommand \BibitemOpen [0]{}%
\providecommand \bibitemStop [0]{}%
\providecommand \bibitemNoStop [0]{.\EOS\space}%
\providecommand \EOS [0]{\spacefactor3000\relax}%
\providecommand \BibitemShut  [1]{\csname bibitem#1\endcsname}%
\let\auto@bib@innerbib\@empty
\end{thebibliography}%






\end{document}